\documentclass[preprint,12pt]{elsarticle}

\usepackage{amsmath,amssymb,amsfonts,amsthm}
\usepackage{mathrsfs}
\usepackage{siunitx}

\usepackage{booktabs}
\usepackage{multirow}
\usepackage{array}
\usepackage{tabularx}
\usepackage{adjustbox}
\usepackage{threeparttable}
\usepackage{rotating}
\usepackage{pdflscape}
\usepackage{longtable}
\usepackage{needspace}
\usepackage{rotating}

\renewcommand{\arraystretch}{1.15}
\usepackage{makecell}
\usepackage{graphicx}
\usepackage{xcolor}
\usepackage{caption}
\usepackage{subcaption}
\usepackage{tikz}
\usepackage{rotating}

\usetikzlibrary{
    arrows.meta,
    positioning,
    calc,
    fit,
    shapes.geometric,
    shapes.misc,
    backgrounds,
    decorations.pathreplacing,
    shadows
}

\usepackage{algorithm}
\usepackage{algpseudocode}
\usepackage{listings}

\usepackage[title]{appendix}
\usepackage{placeins}
\usepackage{textcomp}
\usepackage{manyfoot}
\usepackage{lipsum}

\usepackage{rotating}

\newcommand\MyBox[2]{
    \fbox{\lower0.75cm
        \vbox to 1.3cm{\vfil
            \hbox to 1.3cm{
                \hfil\parbox{0.5cm}{#1}\hfil
            }
            \vfil
        }
    }
}

\usepackage[hidelinks]{hyperref}

\usepackage{tikz}
\usetikzlibrary{arrows.meta, positioning, calc, fit, shapes.geometric}

\begin{document}
\setlength{\emergencystretch}{2em}

\begin{frontmatter}

%% Title, authors, and addresses

%% use the context command for theassociated footnote;

\title{HARP-Net: Hedging-Aware Refinement Pricing Network for Option Pricing}

\author[1,3]{Aryan Bhambu\fnref{equal}}
\ead{a.bhambu@alumni.iitg.ac.in}

\author[2]{Pratyush Baliarsingh\fnref{equal}}
\ead{pratyush.baliarsingh2004@gmail.com}

\author[1]{Koushik Bera}
\ead{koushikbera59@gmail.com}

\author[3]{Selvaraju Natarajan\corref{cor1}}
\ead{nselvaraju@iitg.ac.in}

\address[1]{IITB-Citadel Securities Quantitative Research Lab,
Indian Institute of Technology Bombay, Mumbai 400076, India}

\address[2]{Department of Mathematics,
National Institute of Technology Agartala, Tripura 799046 India}

\address[3]{Department of Mathematics,
Indian Institute of Technology Guwahati,
Guwahati 781039, India}

\address[4]{Department of Mathematics,
Indian Institute of Technology Bombay,
Mumbai 400076, India}

\fntext[equal]{These authors contributed equally to this work
and share first authorship.}

\cortext[cor1]{Corresponding author at: Department of Mathematics,
Indian Institute of Technology Guwahati, Guwahati 781039, India.}

\begin{abstract}

Dynamic option hedging requires a pricing function whose local sensitivities remain reliable throughout complete hedge-to-expiry paths. However, most neural option pricing models are trained as pointwise price predictors, so their induced hedge ratios are only indirectly connected to realized replication performance. We propose the Hedging-Aware Refinement Pricing Network (HARP-Net), a neural option pricing framework that learns an explicit and differentiable pricing surface while directly aligning optimization with pathwise hedging outcomes. HARP-Net combines four components: \emph{(i)} a temporal encoder that summarizes rolling option-state dynamics, \emph{(ii)} domain-wise embeddings that preserve the distinct financial roles of heterogeneous state variables, \emph{(iii)} a bounded residual anchor that stabilizes the initial latent pricing state, and \emph{(iv)} a gated bridge-refinement mechanism that applies controlled context-dependent corrections. The final pricing surface is differentiated with respect to the underlying asset to obtain the model-implied delta, and the network is trained using squared self-financing hedging PnL rather than relying solely on price-regression supervision. We evaluate HARP-Net on four DJX, and XEO call and put datasets against analytical, machine learning, and deep learning baselines. The results show that HARP-Net achieves the strongest overall hedging performance across the reported assessment metrics while maintaining a favorable balance between computational cost and predictive capabilities.

\end{abstract}


\end{frontmatter}

\section{Introduction}
\label{sec:introduction}

Option pricing plays a significant role in quantitative finance by providing a framework for derivative valuation, risk management, portfolio construction, and hedging \cite{black1973pricing,hutchinson1994nonparametric,ruf2019neural}. Accurate estimation of option prices enables investors and financial institutions to evaluate derivative contracts and develop suitable strategies for managing financial risks. Traditional mathematical approaches such as the Black-Scholes model \cite{black1973pricing} have been widely employed for option pricing due to their analytical properties and established financial foundations. Subsequent developments, including the Merton jump-diffusion model \cite{merton1976option} and the Heston stochastic volatility model \cite{heston1993closed}, incorporate discontinuous price movements and time-varying volatility to represent different characteristics of financial markets. Numerical approaches such as the finite element method (FEM) \cite{andalaft2011real} further extend the applicability of mathematical models to pricing problems where closed-form solutions are difficult to obtain. These models provide a theoretical foundation for estimating option prices and their corresponding sensitivities. Recent studies have further investigated the integration of mathematical pricing models with neural networks to retain financial information while learning nonlinear relationships from data \cite{pande2025european,kim2025improved,chen2026learning}.

Machine learning (ML) and deep learning (DL) models have received considerable attention for option pricing due to their ability to capture complex relationships among financial variables \cite{hutchinson1994nonparametric,ruf2019neural}. Early studies employed multilayer perceptrons (MLPs) to approximate nonlinear pricing functions from market observations \cite{hutchinson1994nonparametric,bhambu2025high}. Subsequent research has investigated tree-based ensemble models \cite{bhambu2025randomized}, including Random Forest \cite{breiman2001random,ivascu2021option} and Extreme Gradient Boosting (XGBoost) \cite{chen2016xgboost}, to represent interactions among option pricing variables. Deep learning architectures further extend the representation capabilities of conventional learning approaches through hierarchical nonlinear transformations. For example, residual neural networks (ResNet) incorporate skip connections to facilitate feature learning and have been investigated for option pricing \cite{gan2024option}. Recent studies have also examined the applicability of different neural network architectures and neural stochastic differential equation models for financial derivative pricing \cite{fan2026machine,dellacorte2026machine}. In addition to directly learning option pricing functions, hybrid approaches have been developed to integrate analytical pricing information with data-driven representations. Garcia and Gencay \cite{garcia2000pricing} introduced financial information into neural network-based pricing and hedging through the homogeneity property of option prices. More recently, Kim et al. \cite{kim2025improved} employed residual learning to refine analytical pricing approximations under stochastic volatility models, while Chen et al. \cite{chen2026learning} investigated the estimation of pricing corrections using additional market information. These developments motivate the integration of analytical pricing information with nonlinear feature learning within a unified pricing framework.

The sequential behavior of financial variables provides additional information for modeling the evolution of option prices. In particular, the underlying asset price, moneyness, time to maturity, interest rates, and implied volatility exhibit changes over successive trading periods that can be represented using temporal learning architectures \cite{fritzsch2026predicting}. Recurrent neural networks (RNNs), including Long Short-Term Memory (LSTM) networks \cite{hochreiter1997long} and Gated Recurrent Units (GRUs) \cite{chung2014empirical}, employ recurrent operations to capture dependencies among sequential observations. DeepAR \cite{salinas2020deepar} extends recurrent learning to probabilistic forecasting, while one-dimensional convolutional neural networks (CNN-1D) extract local temporal features from financial observations \cite{liu2023option}. Hybrid architectures such as CNN-LSTM \cite{hu2026multi} and CNN-BiLSTM \cite{sharma2025enhancing} combine convolutional feature extraction with recurrent learning to represent local and sequential dependencies. Recent developments in time series forecasting have introduced different mechanisms for temporal representation, including the linear forecasting architectures DLinear and NLinear \cite{zeng2023transformers}, Transformer-based attention mechanisms \cite{vaswani2017attention}, patch-based attention (PAttn) \cite{tan2024language}, TimeXer \cite{wang2024timexer}, and wavelet-based multiresolution learning through WPMixer \cite{murad2025wpmixer}. Furthermore, recent option pricing studies have investigated hybrid Transformer-LSTM architectures and historical implied volatility observations to account for changing market conditions \cite{xiao2026temporal,fritzsch2026predicting}. Sequential feature learning has also been investigated for estimating option hedging positions directly from historical market data \cite{nian2021learning}. These studies provide a basis for incorporating temporal information into learning frameworks for option pricing and hedging.

In addition to estimating option prices, an important objective of financial modeling is to determine suitable hedge ratios for managing the risks associated with derivative positions. Conventional neural network-based pricing approaches generally minimize the differences between predicted and observed option prices, while hedging performance depends on the sensitivities of the estimated pricing function \cite{garcia2000pricing,ruf2019neural}. Differential machine learning has been introduced to improve the estimation of financial quantities and their derivatives within a common learning framework \cite{huge2020differential,amici2024calibrating}. Recent studies have further investigated generative neural networks and deep backward stochastic differential equations for the joint estimation of option prices and Greeks \cite{cho2025generative,negyesi2026deep}. In parallel, deep hedging approaches directly incorporate financial risk and replication objectives into the learning procedure \cite{buehler2019deep,ruf2022hedging}. Subsequent developments have extended these approaches through residual corrections to analytical hedge ratios, sequential feature learning, implied volatility surface information, and generative market models \cite{qiao2024enhancing,nian2021learning,franccois2025deep,cont2026datadriven}. Recent research has also investigated reinforcement learning for option hedging under different market conditions \cite{chen2026reinforcement,egebjerg2026deep}. Furthermore, differentiable pricing frameworks have been developed to establish a direct relationship between estimated option prices and hedging strategies by incorporating dynamic replication and self-financing conditions into model training \cite{aboussalah2024ai,baradel2025constrained}. These developments motivate the joint consideration of analytical pricing information, temporal feature representation, and hedging performance when constructing a differentiable option pricing model.

Inspired by these developments, this paper proposes a novel Hedging-Aware Refinement Pricing Network (HARP-Net) for option pricing and financial hedging. The proposed model integrates the analytical properties of mathematical pricing models with the nonlinear and temporal feature learning capabilities of deep neural networks. The framework employs a conditional analytical pricing prior constructed from the current option state to provide an initial representation of the pricing information. A temporal encoder is utilized to extract sequential dependencies from a rolling window of historical option-state observations. In addition to temporal feature extraction, the proposed model incorporates domain-wise embeddings to represent different groups of financial variables according to their respective characteristics. These embeddings enable the model to preserve information associated with the underlying asset, option contract, interest rates, and volatility while learning nonlinear relationships among them. The extracted features are subsequently integrated with the analytical pricing information through a bounded residual anchoring mechanism. This mechanism controls the magnitude of the initial corrections and establishes a latent pricing representation for subsequent refinement.

The proposed HARP-Net employs multiple gated bridge-refinement stages to progressively refine the initial latent pricing representation using the extracted temporal and domain-wise features. Each stage generates a nonlinear correction to the representation obtained from the preceding stage, while the gating mechanism controls the contribution of the learned correction. This hierarchical refinement enables the model to incorporate additional financial information while retaining the analytical prior as a reference. The refined representation is subsequently utilized to estimate option prices through an explicit and differentiable pricing function. The corresponding hedge ratios are computed by differentiating the estimated prices with respect to the underlying asset price. A self-financing squared hedging profit-and-loss objective is employed to optimize the network based on the replication errors associated with the estimated hedge ratios. This learning procedure establishes a direct relationship between the refinement of the pricing representation and realized hedging performance. The proposed model is evaluated on four option datasets comprising call and put contracts associated with DJX and XEO. A comprehensive comparative study involving twenty benchmark models is conducted to examine the pricing accuracy and hedging performance of the proposed framework. The main contributions of this study are summarized as follows:

\begin{enumerate}
    \item This study proposes a novel Hedging-Aware Refinement Pricing Network (HARP-Net) that integrates analytical pricing information with temporal feature extraction and domain-wise embeddings to represent sequential and nonlinear relationships among financial variables.

    \item The proposed model introduces a bounded residual anchoring mechanism and multiple gated bridge-refinement stages to progressively refine the latent pricing representation while controlling the magnitude and contribution of the learned corrections.

    \item The proposed framework incorporates a self-financing squared hedging profit-and-loss objective to optimize an explicit, differentiable option-pricing function based on the realized replication errors associated with its estimated hedge ratios.

    \item The computational experiments across four option datasets evaluate the pricing accuracy and hedging performance of the proposed model against twenty benchmark models comprising mathematical pricing approaches, conventional machine learning models, and recent deep learning architectures.
\end{enumerate}

\section{Related Works}
\label{sec:related_work}

This section presents a review of neural network-based option pricing approaches and their developments through analytical pricing information and temporal feature learning. Subsequently, significant developments in deep hedging and differentiable pricing frameworks are discussed.

\subsection{Neural Option Pricing}

Neural networks (NNs) have been widely investigated for option pricing due to their ability to approximate nonlinear relationships among financial variables. Hutchinson et al. \cite{hutchinson1994nonparametric} introduced a nonparametric learning approach for pricing and hedging derivative securities. Subsequent studies have investigated tree-based ensemble models, convolutional networks, and residual architectures to improve the representation of financial information \cite{ivascu2021option,liu2023option,gan2024option}. Tree-based approaches such as Random Forest \cite{breiman2001random} and XGBoost \cite{chen2016xgboost} combine multiple decision trees to represent nonlinear interactions among pricing variables. Recent developments have further examined different neural network architectures and neural stochastic differential equation models for financial derivative pricing \cite{dellacorte2026machine,fan2026machine}. In addition to directly approximating option prices, several studies have investigated the incorporation of analytical pricing information into neural networks. Garcia and Gencay \cite{garcia2000pricing} employed the homogeneity property of option pricing functions to introduce financial structure into neural network-based pricing and hedging. More recently, Pande et al. \cite{pande2025european} introduced a physics-informed residual learning framework for European option pricing under regime-switching dynamics. Kim et al. \cite{kim2025improved} employed residual learning to improve analytical pricing approximations under stochastic volatility models, while Chen et al. \cite{chen2026learning} investigated nonlinear corrections to traditional pricing models using additional market information. These developments support the integration of analytical pricing information with nonlinear feature learning to construct financially informed pricing representations.

The temporal behavior of financial variables provides additional information for modeling the evolution of option prices. Recurrent architectures such as Long Short-Term Memory (LSTM) networks \cite{hochreiter1997long} and Gated Recurrent Units (GRUs) \cite{chung2014empirical} employ sequential state updates to retain information from previous observations. DeepAR \cite{salinas2020deepar} extends recurrent learning to probabilistic time series forecasting, while one-dimensional convolutional neural networks (CNN-1D) employ local filtering operations to extract temporal features \cite{liu2023option}. Hybrid architectures such as CNN-LSTM \cite{hu2026multi} and CNN-BiLSTM \cite{sharma2025enhancing} integrate convolutional feature extraction with recurrent learning to capture local and sequential dependencies. Recent developments have introduced linear forecasting architectures such as DLinear and NLinear \cite{zeng2023transformers}, which employ decomposition and normalization operations, respectively. Transformer \cite{vaswani2017attention} utilizes self-attention mechanisms to represent dependencies among input observations, while PAttn \cite{tan2024language} employs patch-based attention for temporal feature extraction. TimeXer \cite{wang2024timexer} incorporates exogenous variables into the forecasting process, whereas WPMixer \cite{murad2025wpmixer} combines wavelet decomposition with patch-based mixing to represent temporal patterns at different resolutions. More recently, Xiao et al. \cite{xiao2026temporal} investigated a hybrid Transformer-LSTM architecture for option pricing, while Fritzsch et al. \cite{fritzsch2026predicting} examined the contribution of historical observations to option price prediction. These studies highlight the relevance of incorporating temporal information alongside current financial variables and provide a basis for learning sequential pricing representations.

\subsection{Deep Hedging and Residual Hedge Learning}

Hedging plays an important role in financial risk management by determining suitable trading positions to reduce the risks associated with derivative contracts. Traditional hedging strategies utilize sensitivities obtained from analytical pricing functions to determine positions in the underlying asset. Neural network-based approaches provide additional flexibility by learning hedging strategies from financial observations \cite{ruf2019neural}. Buehler et al. \cite{buehler2019deep} introduced a deep hedging framework that incorporates transaction costs, trading constraints, and financial risk measures into a unified learning objective. Subsequent developments have extended deep hedging to quadratic hedging in incomplete markets with jumps \cite{agram2024deep}, end-to-end option trading \cite{tan2024deep}, and efficient training through second-order optimization \cite{mueller2024fast}. Ruf and Wang \cite{ruf2022hedging} further investigated neural networks and regression-based approaches for estimating hedge ratios by directly minimizing hedging errors. These developments establish the importance of incorporating financial risk objectives into the learning procedure to account for the replication performance of estimated hedge ratios.

In addition to directly learning hedging strategies, several studies have investigated the integration of temporal information and analytical hedge ratios into neural network architectures. Nian et al. \cite{nian2021learning} employed recurrent neural networks to learn sequential hedging strategies from historical market observations, enabling the model to account for changes in financial variables over successive trading periods. Qiao et al. \cite{qiao2024enhancing} introduced a residual learning approach to estimate corrections to hedge ratios obtained from the Black-Scholes model. This approach combines the financial information provided by analytical hedge ratios with nonlinear corrections learned from market observations. Francois et al. \cite{franccois2025deep} further investigated deep hedging using implied volatility surface information to represent changing market conditions and volatility risk. These studies provide a basis for combining analytical financial information with temporal feature extraction and residual learning within a common hedging framework.

Recent developments have further examined alternative learning mechanisms for estimating hedging strategies under different market conditions. Cont and Vuletic \cite{cont2026datadriven} introduced a data-driven hedging framework that employs generative models to represent market scenarios and determine hedging positions through conditional risk minimization. Chen et al. \cite{chen2026reinforcement} investigated the relationship between implied volatility fitting and realized hedging performance using reinforcement learning with financial shortfall objectives. Egebjerg \cite{egebjerg2026deep} further examined the applicability of deep hedging to options with short times to maturity. These developments highlight the relevance of learning objectives based on realized hedging performance. In particular, the integration of analytical hedging information with nonlinear and temporal feature learning motivates the construction of differentiable pricing models that incorporate replication objectives directly into the learning procedure.

\subsection{Hedging-Aware Differentiable Pricing}

Differentiable neural networks have received considerable attention for the joint estimation of financial quantities and their sensitivities. Differential machine learning \cite{huge2020differential} incorporates function values and their derivatives into the learning procedure to improve the approximation of financial relationships. Amici et al. \cite{amici2024calibrating} extended this direction by employing deep differential networks for gradient-based calibration of the Heston model. Neural stochastic differential equation and partial differential equation-based approaches have also been investigated for option pricing under complex financial dynamics \cite{teng2024neural,papapantoleon2025time,fan2026machine}. Recent studies have further examined approximation errors associated with deep partial differential equation solvers \cite{rou2025error} and the incorporation of market-implied volatility surfaces into option pricing frameworks \cite{ding2025deep}. These developments demonstrate the applicability of differentiable neural representations for integrating financial information into option pricing and sensitivity estimation.

Subsequent studies have investigated the simultaneous estimation of option prices and their corresponding Greeks using different neural learning mechanisms. Cho et al. \cite{cho2025generative} introduced a generative neural network framework based on conditional normalizing flows for estimating option prices and Greeks. Guo et al. \cite{guo2025simultaneous} investigated neural network approaches for obtaining upper and lower pricing bounds of American-style options together with their hedging strategies. Negyesi and Oosterlee \cite{negyesi2026deep} employed deep backward stochastic differential equations for simultaneous pricing and delta-gamma hedging of high-dimensional Bermudan option portfolios. Sakuma \cite{sakuma2026differential} further investigated differential machine learning for estimating prices and Greeks of zero-days-to-expiry options under stochastic volatility and jumps. Recent derivative-informed operator learning frameworks have also considered the joint representation of financial quantities and their sensitivities \cite{alonso2026derivative}. These developments highlight the importance of retaining an explicit and differentiable pricing representation to establish a relationship between estimated option prices and their corresponding hedge ratios.

In addition to incorporating derivative information into neural network training, recent studies have investigated the integration of financial replication principles directly into the learning procedure. Aboussalah et al. \cite{aboussalah2024ai} introduced Finance-Informed Neural Networks (FINN) to learn option pricing functions using dynamic hedging principles. Baradel \cite{baradel2025constrained} developed a constrained deep learning framework for European option pricing and hedging in incomplete markets, where the derivatives of the estimated pricing function determine the corresponding hedge ratios. These approaches provide a methodological foundation for connecting differentiable pricing functions with self-financing replication objectives. Together with recent developments in analytical residual learning \cite{pande2025european,kim2025improved,chen2026learning} and sequential financial modeling \cite{nian2021learning,xiao2026temporal,fritzsch2026predicting}, these studies motivate the proposed HARP-Net framework. The proposed model extends this direction by integrating conditional analytical pricing priors, temporal feature extraction, domain-wise embeddings, bounded residual anchoring, and gated hierarchical refinement within a unified hedging-aware pricing architecture.

\section{Proposed Methodology}
\label{sec:proposed_methodology}

\begin{sidewaysfigure}[!p]
    \centering
    \resizebox{\textheight}{!}{%
        \input{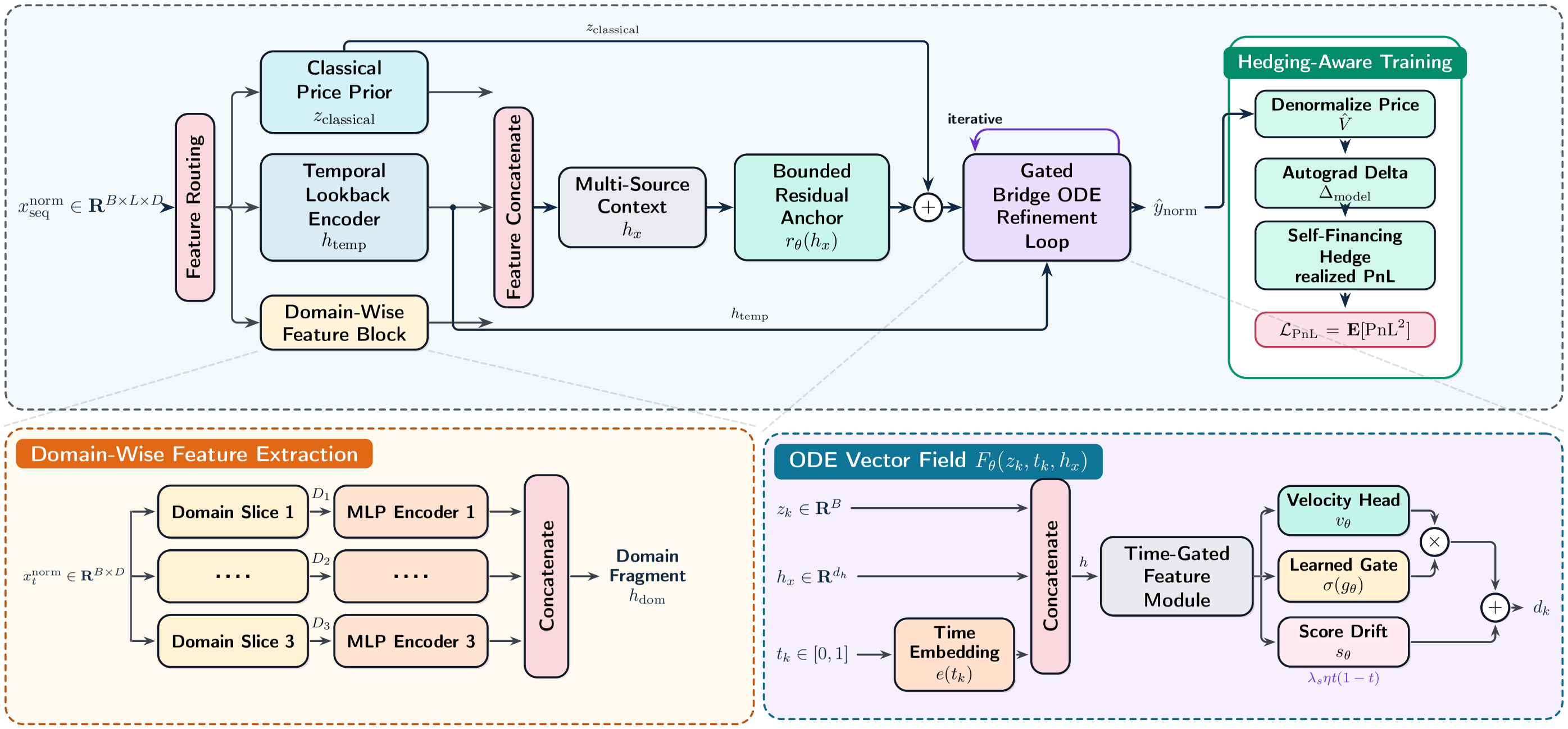}%
    }
    \caption{Architecture of the proposed HARP-Net.}
    \label{fig:harpnet_architecture}
\end{sidewaysfigure}

This paper proposes a novel Hedging-Aware Refinement Pricing Network (HARP-Net) for option pricing and financial hedging. The proposed model integrates analytical pricing information with the nonlinear and temporal feature learning capabilities of deep neural networks. An analytical pricing prior is initially obtained from the current financial state, which is subsequently refined using the features extracted from historical and current observations. The proposed architecture comprises four major components: a temporal context encoder, a domain-wise embedding block, a bounded residual anchor, and a gated bridge-refinement mechanism. The resulting pricing function is differentiated to obtain the corresponding hedge ratios, while the network parameters are optimized using realized hedging errors. The architecture of the proposed HARP-Net is illustrated in Fig.~\ref{fig:harpnet_architecture}.

\subsection{Problem Formulation}
\label{subsec:analytical_prior}

Let $x_t$ denote the financial state of a European option at time $t$, represented as
\begin{equation}
    x_t=
    [m_t,k_t,\tau_t,r_t,\nu_t,q_t]^\top
    \in\mathbb{R}^{D},
    \label{eq:methodology_state}
\end{equation}
where $m_t=\log(S_t/K)$ and $k_t=\log K$ denote the log-moneyness and logarithmic strike price, respectively. Here, $S_t$ and $K$ represent the underlying asset price and strike price, while $\tau_t$, $r_t$, $\nu_t$, and $q_t$ denote the time to maturity, risk-free rate, implied volatility, and dividend yield proxy, respectively. The normalized rolling window of financial observations is represented as
\begin{equation}
    \widetilde X_t^{(L)}
    =
    \begin{bmatrix}
        \widetilde x_{t-L+1}^{\top}\\
        \vdots\\
        \widetilde x_t^{\top}
    \end{bmatrix}
    \in\mathbb{R}^{L\times D},
    \label{eq:methodology_input}
\end{equation}
where $L$ denotes the lookback length, $D=6$ represents the number of input variables, and $\widetilde x_t$ denotes the normalized financial state obtained using the preprocessing procedure described in Section~\ref{subsec:data_preprocessing}. For a batch of $B$ input windows, the resulting input dimension is $B\times L\times D$. The proposed model learns a differentiable pricing function $\mathcal{F}_{\Theta}:\mathbb{R}^{L\times D}\rightarrow\mathbb{R}$, where the unconstrained normalized option price is obtained as
\begin{equation}
    \widehat y_t^{\mathrm{norm}}
    =
    \mathcal{F}_{\Theta}(\widetilde X_t^{(L)}).
\end{equation}

HARP-Net employs a dividend-adjusted Black--Scholes pricing function \cite{black1973pricing} to construct an initial pricing representation. Since the analytical pricing function operates on the original financial variables, the normalized terminal observation is transformed back to its original scale as
\begin{equation}
\begin{aligned}
    x_t &= \widetilde x_t\odot a_x+b_x,\\
    K &= e^{k_t},\qquad S_t=e^{k_t+m_t},
\end{aligned}
    \label{eq:methodology_inverse_normalization}
\end{equation}
where $a_x$ and $b_x$ denote the inverse normalization coefficients and $\odot$ represents element-wise multiplication. The dividend-adjusted Black--Scholes quantities are computed as
\begin{equation}
\begin{aligned}
    d_{1,t}
    &=
    \frac{
    m_t+(r_t-q_t+\nu_t^2/2)\tau_t
    }{\nu_t\sqrt{\tau_t}},\\
    d_{2,t}
    &=d_{1,t}-\nu_t\sqrt{\tau_t}.
\end{aligned}
    \label{eq:prior_d1}
\end{equation}
Depending on the option type $a\in\{C,P\}$, the corresponding analytical price is obtained as
\begin{equation}
    V_t^{\mathrm{cp}}
    =
    \begin{cases}
        S_te^{-q_t\tau_t}\Phi(d_{1,t})
        -Ke^{-r_t\tau_t}\Phi(d_{2,t}), & a=C,\\[3pt]
        Ke^{-r_t\tau_t}\Phi(-d_{2,t})
        -S_te^{-q_t\tau_t}\Phi(-d_{1,t}), & a=P,
    \end{cases}
    \label{eq:classical_price_prior}
\end{equation}
where $\Phi(\cdot)$ denotes the cumulative distribution function of the standard normal distribution. The obtained analytical price is subsequently transformed into the normalized output space as
\begin{equation}
    z_t^{\mathrm{cp}}
    =
    \frac{V_t^{\mathrm{cp}}-b_y}{a_y},
    \label{eq:normalized_classical_prior}
\end{equation}
where $a_y$ and $b_y$ denote the inverse normalization coefficients associated with option premiums. Numerical floors and exponent clipping are employed to avoid invalid divisions and overflow, with the preceding expressions representing the prior when these safeguards are inactive. The analytical prior is computed within the computational graph to retain its dependence on the underlying asset price during differentiation. The resulting representation serves as an initial pricing state for the subsequent feature learning and refinement operations.

\subsection{Temporal and Domain-Wise Feature Learning}
\label{subsec:temporal_encoder}
\label{subsec:domain_embedding}

The analytical prior obtained from the current financial state does not explicitly consider preceding market observations. To incorporate temporal dependencies, HARP-Net employs a single-layer unidirectional GRU \cite{chung2014empirical} that processes the normalized rolling window $\widetilde X_t^{(L)}$. Let $u_{t,\ell}$ denote the recurrent hidden state corresponding to the $\ell$-th observation of the input window. The temporal representation is obtained as
\begin{equation}
\begin{aligned}
    u_{t,0}&=0,\\
    u_{t,\ell}
    &=
    \operatorname{GRU}_{\alpha}
    (u_{t,\ell-1},\widetilde x_{t-L+\ell}),\\
    h_t^{\mathrm{temp}}
    &=u_{t,L}\in\mathbb{R}^{d},
\end{aligned}
    \label{eq:temporal_representation}
\end{equation}
where $\ell=1,\ldots,L$, $\alpha$ denotes the trainable GRU parameters, and $d$ represents the latent dimension. The hidden state is initialized to zero for each input window, and no dropout is employed within the GRU.

In addition to temporal information, the current financial state comprises variables with different financial characteristics. The proposed model independently processes these variables through three domain-wise embeddings representing the contract state, maturity and carry, and implied volatility. The corresponding feature representations are obtained as
\begin{equation}
\begin{aligned}
    e_t^{(1)}
    &=\phi_1([\widetilde m_t,\widetilde k_t]^\top),\\
    e_t^{(2)}
    &=\phi_2([\widetilde\tau_t,
    \widetilde r_t,\widetilde q_t]^\top),\\
    e_t^{(3)}
    &=\phi_3(\widetilde\nu_t),
\end{aligned}
\end{equation}
where $\phi_1$, $\phi_2$, and $\phi_3$ denote independent multilayer perceptrons comprising hidden layers of width $d$ with nonlinear activation and dropout, followed by a linear output layer. Each domain-wise embedding has dimension $d$. The obtained features are concatenated as
\begin{equation}
    e_t^{\mathrm{dom}}
    =
    [e_t^{(1)};e_t^{(2)};e_t^{(3)}]
    \in\mathbb{R}^{3d},
    \label{eq:domain_concatenation}
\end{equation}
where $[\,;\,]$ represents vector concatenation. This enables the model to learn separate nonlinear representations of different financial variables before combining their features.

The domain-wise features are subsequently concatenated with the normalized current state, temporal representation, and analytical prior as
\begin{equation}
    v_t=
    [e_t^{\mathrm{dom}};
    \widetilde x_t;
    h_t^{\mathrm{temp}};
    z_t^{\mathrm{cp}}]
    \in\mathbb{R}^{4d+7}.
    \label{eq:context_concatenation}
\end{equation}
The resulting features are transformed into a common latent context as
\begin{equation}
\begin{aligned}
    c_t=\operatorname{LayerNorm}\Big(
    &W_2\mathcal{D}_p\big(
    \sigma(W_1v_t+b_1)
    \big)+b_2\Big)
    \in\mathbb{R}^{d},
\end{aligned}
    \label{eq:context_fusion}
\end{equation}
where $\sigma(\cdot)$ denotes the activation function, $\mathcal{D}_p(\cdot)$ represents dropout with probability $p$, and $W_1\in\mathbb{R}^{d\times(4d+7)}$ and $W_2\in\mathbb{R}^{d\times d}$ denote the corresponding weight matrices. The inclusion of the original financial variables and analytical prior preserves their information alongside the extracted nonlinear features. The resulting contextual representation is utilized in the subsequent residual anchoring and bridge-refinement operations.

\subsection{Bounded-Residual Anchor}
\label{subsec:residual_anchor}

The contextual representation obtained from the preceding components is utilized to refine the analytical prior through a bounded residual learning mechanism \cite{kim2025improved,pande2025european}. Instead of directly estimating an initial option price, HARP-Net learns an additive correction to the analytical prior as
\begin{align}
    \rho_t
    &=
    \operatorname{clip}\Big(
    \lambda_{\mathrm{cp}}\lambda_r
    \tanh(w_r^\top c_t+b_r),
    -c_r,c_r
    \Big),
    \label{eq:bounded_residual}\\
    y_{t,0}
    &=z_t^{\mathrm{cp}}+\rho_t,
    \label{eq:residual_anchor_state}
\end{align}
where $w_r$ and $b_r$ denote the trainable residual parameters, $\lambda_{\mathrm{cp}}$ and $\lambda_r$ represent the nonnegative residual scaling coefficients, and $c_r>0$ denotes the optional clipping threshold. Since the output of the $\tanh(\cdot)$ activation function is bounded, the initial correction satisfies
\begin{equation}
    |y_{t,0}-z_t^{\mathrm{cp}}|
    \leq
    \min(\lambda_{\mathrm{cp}}\lambda_r,c_r).
    \label{eq:residual_anchor_bound}
\end{equation}
The corresponding bound without clipping is given by $\lambda_{\mathrm{cp}}\lambda_r$.

The proposed model employs $c_r=2$, while the product of the residual scaling coefficients does not exceed $0.5$, making explicit residual clipping inactive under the considered configurations. Furthermore, the residual output layer is initialized with zero weights and bias such that $y_{t,0}=z_t^{\mathrm{cp}}$ before training. Thus, the bounded residual anchor retains the analytical prior as an initial pricing representation while allowing the network to learn contextual corrections for subsequent bridge refinement.

\subsection{Gated Bridge-Refinement}
\label{subsec:bridge_refinement}

The initial pricing state obtained from the residual anchor is progressively refined through a deterministic bridge-inspired vector field. Let $s\in[0,1]$ denote the internal refinement coordinate, independent of the financial observation time and option maturity. The corresponding continuous formulation is expressed as
\begin{equation}
    \frac{\mathrm{d}y_t(s)}{\mathrm{d}s}
    =
    \mathcal{G}_{\theta}(c_t,y_t(s),s),
    \qquad y_t(0)=y_{t,0},
    \label{eq:bridge_ode}
\end{equation}
where $\mathcal{G}_{\theta}$ denotes the conditional vector field. The refinement interval is divided into $J$ Euler-type steps with midpoint time evaluation, where $\Delta s=1/J$ and $s_k=(k+1/2)\Delta s$ for $k=0,\ldots,J-1$. To incorporate the refinement time, sinusoidal features and their corresponding nonlinear projection are obtained as
\begin{align}
    \zeta(s)
    &=
    [\sin(s\boldsymbol{\omega});
    \cos(s\boldsymbol{\omega});s;1-s],\\
    e(s)
    &=
    W_{e,2}\operatorname{Softplus}
    (W_{e,1}\zeta(s)+b_{e,1})
    +b_{e,2},
    \label{eq:bridge_time_embedding}
\end{align}
where $e(s)\in\mathbb{R}^{d}$ and $\omega_i=10000^{-i/(H-1)}$ for $i=0,\ldots,H-1$, with $H=\lfloor d/2\rfloor>1$. The time embedding is combined with the contextual representation and current pricing state as
\begin{equation}
    h_{t,k}^{(0)}
    =
    W_{\mathrm{in}}[c_t;e(s_k);y_{t,k}]
    +b_{\mathrm{in}},
    \label{eq:bridge_input_projection}
\end{equation}
where $W_{\mathrm{in}}\in\mathbb{R}^{d\times(2d+1)}$ and $b_{\mathrm{in}}\in\mathbb{R}^{d}$ denote the projection parameters.

The resulting representation is processed through $H_f$ time-modulated residual blocks. The modulation vector and hidden representation corresponding to the $\ell$-th block are computed as
\begin{align}
    a_k^{(\ell)}
    &=
    \operatorname{Softplus}
    (W_{g,\ell}e(s_k)+b_{g,\ell}),
    \label{eq:bridge_time_modulation}\\
    h_{t,k}^{(\ell)}
    &=
    h_{t,k}^{(\ell-1)}
    +\mathcal{D}_p\Big(
    \sigma\big(
    (W_\ell h_{t,k}^{(\ell-1)}+b_\ell)
    \odot a_k^{(\ell)}
    \big)\Big),
    \label{eq:bridge_residual_block}
\end{align}
where $\ell=1,\ldots,H_f$ and the hidden representations and modulation vectors have dimension $d$. The time-dependent modulation adjusts the contribution of latent features at different refinement steps, while the residual connections retain the preceding representations. The output of the final residual block is utilized to generate a scalar velocity and its corresponding gate as
\begin{equation}
\begin{aligned}
    v_{t,k}
    &=w_v^\top h_{t,k}^{(H_f)}+b_v,\\
    \gamma_{t,k}
    &=
    \operatorname{sigmoid}
    (w_g^\top h_{t,k}^{(H_f)}+b_g).
\end{aligned}
    \label{eq:bridge_velocity_gate}
\end{equation}
The refinement drift and subsequent pricing state are obtained as
\begin{align}
    d_{t,k}
    &=
    \operatorname{clip}
    (\gamma_{t,k}v_{t,k},-c_d,c_d),
    \label{eq:bridge_clipped_drift}\\
    y_{t,k+1}
    &=y_{t,k}+\Delta s\,d_{t,k},
    \qquad
    \widehat y_t^{\mathrm{norm}}=y_{t,J},
    \label{eq:bridge_state_update}
\end{align}
where $c_d>0$ denotes the drift clipping threshold. The gating operation controls the contribution of the velocity term, while clipping restricts the magnitude of each correction. The same vector-field parameters are employed across all refinement steps. The optional score-based correction and bridge-matching objective remain inactive in the primary experiments.

Since $|d_{t,k}|\leq c_d$ and $J\Delta s=1$, the cumulative correction satisfies
\begin{equation}
    |y_{t,J}-y_{t,0}|
    \leq
    \frac{1}{J}\sum_{k=0}^{J-1}|d_{t,k}|
    \leq c_d.
    \label{eq:cumulative_bridge_bound}
\end{equation}
Combining this inequality with Eq.~\eqref{eq:residual_anchor_bound}, the overall deviation from the analytical prior is bounded as
\begin{equation}
    |\widehat y_t^{\mathrm{norm}}-z_t^{\mathrm{cp}}|
    \leq
    \min(\lambda_{\mathrm{cp}}\lambda_r,c_r)+c_d.
    \label{eq:total_prior_deviation_bound}
\end{equation}
Thus, the proposed refinement mechanism controls the overall correction in the normalized pricing space independently of the number of refinement steps, although it does not directly constrain the pricing derivatives. The velocity output layer is initialized with zero weights and bias, while the gating layer is initialized with zero weights and a bias of $-2$. Together with the zero-initialized residual anchor, the initial backbone output corresponds to the analytical prior before applying the price-interval transformation.

\subsection{Hedging-Aware Training}
\label{subsec:hedging_training}

The proposed HARP-Net employs a hedging-based learning objective to optimize the pricing function according to realized replication errors \cite{buehler2019deep,aboussalah2024ai,baradel2025constrained}. The output obtained from the final refinement step is inverse-normalized and subjected to the common European price-interval transformation $\mathcal{B}_{\varepsilon}$ described in Section~\ref{subsec:experimental_setup}. The constrained option price and its normalized representation are obtained as
\begin{equation}
\begin{aligned}
    \widehat V_t
    &=
    \mathcal{B}_{\varepsilon}
    (a_y y_{t,J}+b_y;S_t,K,\tau_t,r_t,q_t),\\
    \widehat z_t
    &=
    \frac{\widehat V_t-b_y}{a_y}.
\end{aligned}
    \label{eq:constrained_option_price}
\end{equation}
Since $m_t=a_m\widetilde m_t+b_m=\log(S_t/K)$, the model-implied delta is obtained using the chain rule as
\begin{equation}
\begin{aligned}
    \Delta_t^{\Theta}
    &=
    \frac{\partial\widehat V_t}{\partial S_t}
    =
    \frac{1}{S_t}
    \frac{\partial\widehat V_t}{\partial m_t}\\
    &=
    \frac{a_y}{S_ta_m}
    \frac{\partial\widehat z_t}
    {\partial\widetilde m_t},
\end{aligned}
    \label{eq:model_implied_delta}
\end{equation}
where $a_m$ and $b_m$ denote the inverse normalization coefficients associated with log-moneyness. The remaining financial variables and historical observations are kept unchanged during differentiation. Hence, the estimated hedge ratio incorporates the dependence of the analytical prior, learned corrections, and price bounds on the underlying asset price.

A self-financing hedging portfolio is constructed for each complete option path with observations $t=0,\ldots,T$, where $0$ and $T$ represent hedge initiation and settlement, respectively. The portfolio consists of a short option position, underlying holdings $\xi_t$, and a cash account $C_t$. The initial portfolio is constructed using the observed option premium $V_0^{\mathrm{obs}}$ as
\begin{equation}
    \xi_0=\Delta_0^{\Theta},
    \qquad
    C_0=V_0^{\mathrm{obs}}-\xi_0S_0.
    \label{eq:initial_cash_account}
\end{equation}
Let $\delta_t$ denote the elapsed time in years between successive observations under the ACT/365F convention. The cash balance before rebalancing is computed as
\begin{equation}
    C_t^-
    =
    C_{t-1}e^{r_{t-1}\delta_t}
    +
    \xi_{t-1}S_{t-1}q_{t-1}\delta_t,
    \label{eq:cash_accrual}
\end{equation}
for $t=1,\ldots,T$, where the two terms represent the cash balance after interest accrual and approximate dividend income, respectively. At each live observation, the portfolio is rebalanced under the zero-transaction-cost setting as
\begin{equation}
\begin{aligned}
    \xi_t&=\Delta_t^{\Theta},\\
    C_t&=C_t^--(\xi_t-\xi_{t-1})S_t,
\end{aligned}
    \label{eq:self_financing_rebalance}
\end{equation}
where $t=1,\ldots,T-1$. At settlement, no additional hedge position is established and the terminal payoff is computed as
\begin{equation}
    \Pi_a(S_T^\star,K)
    =
    \begin{cases}
        \max(S_T^\star-K,0), & a=C,\\
        \max(K-S_T^\star,0), & a=P,
    \end{cases}
    \label{eq:terminal_payoff}
\end{equation}
where $S_T^\star$ denotes the prescribed settlement level. The terminal hedging profit and loss (PnL) associated with the $i$-th option path is obtained as
\begin{equation}
    P_i(\Theta)
    =
    C_T^-+\xi_{T-1}S_T^\star
    -\Pi_a(S_T^\star,K).
    \label{eq:terminal_hedging_pnl}
\end{equation}
The terminal portfolio value and option payoff are evaluated using the same settlement level. The corresponding pathwise hedging procedure is summarized in Algorithm~\ref{alg:pathwise_evaluation}.

\begin{algorithm}[!htbp]
\caption{Pathwise Delta-Hedging Evaluation}
\label{alg:pathwise_evaluation}
\small
\begin{algorithmic}[1]

\Require Option path $p$, pricing model $\mathcal{F}_{\Theta}$,
option type $a$, training-fitted scalers, graph-retention
indicator $\chi$.

\Ensure Terminal hedging PnL $P_{\Theta}(p)$.

\For{$t=0,\ldots,T-1$}
    \State Obtain $\widetilde X_t^{(L)}$ and compute
    $\widehat y_t^{\mathrm{norm}}$.
    \State Compute $\widehat z_t$ using
    Eq.~\eqref{eq:constrained_option_price}.
    \State Compute $\Delta_t^{\Theta}$ using
    Eq.~\eqref{eq:model_implied_delta}, retaining the
    derivative graph if $\chi=1$.
\EndFor

\State Initialize $\xi_0$ and $C_0$ using
Eq.~\eqref{eq:initial_cash_account}.

\For{$t=1,\ldots,T-1$}
    \State Compute $C_t^-$ using Eq.~\eqref{eq:cash_accrual}.
    \State Update $\xi_t$ and $C_t$ using
    Eq.~\eqref{eq:self_financing_rebalance}.
\EndFor

\State Compute final cash accrual $C_T^-$ using
Eq.~\eqref{eq:cash_accrual}.

\State Obtain terminal payoff $\Pi_a(S_T^\star,K)$
using Eq.~\eqref{eq:terminal_payoff}.

\State Compute $P_{\Theta}(p)$ using
Eq.~\eqref{eq:terminal_hedging_pnl}.

\State \Return $P_{\Theta}(p)$.

\end{algorithmic}
\end{algorithm}

The proposed model is trained by minimizing the squared terminal hedging PnL across complete option paths. For $N$ training paths, the hedging loss and its corresponding parameter gradient are expressed as
\begin{align}
    \mathcal{L}_{\mathrm{hedge}}(\Theta)
    &=
    \frac{1}{N}\sum_{i=1}^{N}P_i(\Theta)^2,
    \label{eq:quadratic_hedging_objective}\\
    \nabla_{\Theta}
    \mathcal{L}_{\mathrm{hedge}}(\Theta)
    &=
    \frac{2}{N}
    \sum_{i=1}^{N}
    P_i(\Theta)\nabla_{\Theta}P_i(\Theta).
    \label{eq:hedging_loss_gradient}
\end{align}
Since the initial premium and realized financial variables are independent of the network parameters, the parameter dependence of the hedging loss is introduced through the estimated hedge ratios. Consequently, the optimization procedure requires the mixed derivatives
\begin{equation}
    \nabla_{\Theta}\Delta_t^{\Theta}
    =
    \frac{a_y}{S_ta_m}
    \nabla_{\Theta}
    \left(
    \frac{\partial\widehat z_t}
    {\partial\widetilde m_t}
    \right),
    \label{eq:mixed_hedging_derivatives}
\end{equation}
which are obtained by retaining the derivative graph during training. The training paths are divided into mini-batches of complete option paths, and the network parameters are optimized using Adam. The model associated with the lowest validation hedging loss is retained for subsequent evaluation. Since the portfolio is initialized using the observed premium, the learning objective directly optimizes the hedge sensitivities rather than the absolute price levels. The analytical prior, bounded corrections, and price-interval transformation provide structure to the learned pricing function, although hedging errors alone do not uniquely determine its price level. The training procedure of HARP-Net is summarized in Algorithm~\ref{alg:hedge_loss_training}.

\begin{algorithm}[!htbp]
\caption{Training Procedure of the Proposed HARP-Net}
\label{alg:hedge_loss_training}
\small
\begin{algorithmic}[1]

\Require Training and validation paths
$\mathcal{D}_{\mathrm{train}}$, $\mathcal{D}_{\mathrm{val}}$,
initialized parameters $\Theta$, learning rate $\eta$,
path-batch size $B$, maximum epochs $E$,
patience $p_{\mathrm{stop}}$.

\Ensure Optimized model parameters $\Theta^\star$.

\State Initialize
$\mathcal{L}_{\mathrm{best}}\gets\infty$,
$n_{\mathrm{wait}}\gets0$.

\For{$e=1,\ldots,E$}

    \For{each $\mathcal{B}_j$}
        \State Obtain $P_{\Theta}(p)$ for each
        $p\in\mathcal{B}_j$ using
        Algorithm~\ref{alg:pathwise_evaluation} with $\chi=1$.
        \State Compute
        $\mathcal{L}_j=
        |\mathcal{B}_j|^{-1}
        \sum_{p\in\mathcal{B}_j}P_{\Theta}(p)^2$.
        \State Backpropagate $\mathcal{L}_j$; terminate
        if the loss or gradients are non-finite.
        \State Clip the gradient norm and
        update $\Theta$ using Adam.
    \EndFor

    \State Compute
    $\mathcal{L}_{\mathrm{val}}$ over
    $\mathcal{D}_{\mathrm{val}}$ using
    Algorithm~\ref{alg:pathwise_evaluation} with $\chi=0$
    and Eq.~\eqref{eq:quadratic_hedging_objective}.

    \If{$\mathcal{L}_{\mathrm{val}}<\mathcal{L}_{\mathrm{best}}$}
        \State Save $\Theta^\star\gets\Theta$,
        $\mathcal{L}_{\mathrm{best}}\gets\mathcal{L}_{\mathrm{val}}$,
        $n_{\mathrm{wait}}\gets0$.
    \Else
        \State $n_{\mathrm{wait}}\gets n_{\mathrm{wait}}+1$.
    \EndIf

    \If{$n_{\mathrm{wait}}\geq p_{\mathrm{stop}}$}
        \State \textbf{break}.
    \EndIf

\EndFor

\State Restore and \Return $\Theta^\star$.

\end{algorithmic}
\end{algorithm}

\newcommand{\best}[1]{\textbf{#1}}

\section{Experimental Study}
\label{sec:experimental_study}

This section presents the experimental analysis of the proposed HARP-Net for option pricing and financial hedging. The performance of the proposed model is evaluated across four option datasets and compared with traditional pricing approaches, machine learning models, and recent deep learning architectures.

\subsection{Data Description and Preprocessing}
\label{subsec:data_description}
\label{subsec:data_preprocessing}

This paper utilizes four option datasets corresponding to two widely followed indices in the United States financial market: the Dow Jones Industrial Average (DJX) and the S\&P 100 Index (XEO). Both call and put options associated with these indices are considered for the experimental analysis. The DJX option datasets cover the period from [start date] to [end date], while the XEO option datasets cover the period from [start date] to [end date]. The historical option market data are obtained from IVolatility\footnote{\url{https://www.ivolatility.com/}} using the corresponding option symbols. The datasets comprise observed option premiums, underlying asset prices, strike prices, time to maturity, implied volatility, risk-free rates, and dividend-related information.

The datasets contain option contracts with different strike prices and maturity periods. Based on their moneyness at the first recorded observation, the contracts are categorized into at-the-money (ATM), in-the-money (ITM), and out-of-the-money (OTM) options. A contract is considered ATM if its moneyness lies within the interval $[0.96,1.04]$, whereas contracts outside this interval are classified as ITM or OTM depending on the option type. Table~\ref{tab:dataset_description} presents the number of observations, distinct contracts, and their distribution across the three moneyness categories for each dataset.

\begin{table}[!htbp]
    \centering
    \caption{Description of the option datasets utilized
    for the experimental analysis.}
    \label{tab:dataset_description}
    \small
    \setlength{\tabcolsep}{6pt}
    \renewcommand{\arraystretch}{1.15}

    \begin{threeparttable}
        \begin{tabular}{@{}lrrrrr@{}}
            \toprule
            \textbf{Dataset}
            & \textbf{Observations}
            & \textbf{Contracts}
            & \textbf{ATM}
            & \textbf{ITM}
            & \textbf{OTM} \\
            \midrule
            DJX-Call & 22,038  & 500   & 253   & 122   & 125 \\
            DJX-Put  & 23,284  & 521   & 254   & 142   & 125 \\
            XEO-Call & 122,349 & 3,596 & 1,855 & 1,021 & 720 \\
            XEO-Put  & 138,981 & 4,224 & 1,842 & 1,511 & 871 \\
            \bottomrule
        \end{tabular}

        \begin{tablenotes}[flushleft]
            \footnotesize
            \item \textit{Note:} ATM, ITM, and OTM denote
            at-the-money, in-the-money, and out-of-the-money
            contracts, respectively. The classification is
            based on the moneyness ($S/K$) at the first
            recorded observation of each contract.
        \end{tablenotes}
    \end{threeparttable}
\end{table}

The observations associated with each option contract are arranged chronologically to construct individual option paths. The underlying asset price and strike price are used to obtain two financial variables, namely log-moneyness ($m_t$) and logarithmic strike price ($k_t$), which are defined as
\begin{equation}
    m_t=\log\left(\frac{S_t}{K}\right),
    \qquad
    k_t=\log K,
    \label{eq:preprocessing_log_features}
\end{equation}
where $S_t$ denotes the underlying asset price at time $t$ and $K$ represents the strike price of the corresponding option contract. These variables are combined with the time to maturity, risk-free rate, implied volatility, and dividend yield proxy to construct the six-dimensional financial state defined in Eq.~\eqref{eq:methodology_state}.

Each dataset is divided into training, validation, and testing subsets at the contract level, such that observations associated with an individual contract belong to only one subset. MinMax normalization is employed to preprocess the financial variables and option premiums. Let $a_x$ and $b_x$ denote the feature-wise scaling and shifting coefficients, respectively, and let $a_y$ and $b_y$ represent the corresponding coefficients for the option premiums. The normalized input variables and option premiums are obtained as
\begin{align}
    \widetilde x_t
    &=(x_t-b_x)\oslash a_x,
    \label{eq:input_normalization}\\
    \widetilde V_t^{\mathrm{obs}}
    &=\frac{V_t^{\mathrm{obs}}-b_y}{a_y},
    \label{eq:output_normalization}
\end{align}
where $\oslash$ denotes element-wise division and $V_t^{\mathrm{obs}}$ represents the observed option premium. The normalization coefficients are obtained from the training dataset and subsequently utilized to transform the validation and testing datasets.

The proposed model utilizes a lookback period of $L=20$ observations to incorporate temporal information into the pricing function. The rolling windows are constructed independently for each option contract while preserving the chronological ordering of observations. Although the normalized variables are used for model training, the original financial variables are retained to evaluate the estimated option prices and their corresponding hedging performance.

\subsection{Evaluation Metrics}
\label{subsec:evaluation_metrics}

The performance of the comparative models is evaluated by considering the accuracy of the estimated option prices and the replication performance of their corresponding hedge ratios. The root mean squared error (RMSE) and mean absolute error (MAE) are employed to evaluate the differences between the estimated and observed option premiums. These metrics are expressed as
\begin{align}
    \mathrm{RMSE}_{\mathrm{price}}
    &=
    \sqrt{
    \frac{1}{n}
    \sum_{i=1}^{n}
    (V_i^{\mathrm{obs}}-\widehat V_i)^2
    },
    \label{eq:pricing_rmse}\\
    \mathrm{MAE}_{\mathrm{price}}
    &=
    \frac{1}{n}
    \sum_{i=1}^{n}
    |V_i^{\mathrm{obs}}-\widehat V_i|,
    \label{eq:pricing_mae}
\end{align}
where $V_i^{\mathrm{obs}}$ and $\widehat V_i$ denote the observed and estimated option premiums, respectively, and $n$ represents the number of evaluated observations.

In addition to pricing accuracy, the hedging performance is evaluated using the terminal PnL obtained from the self-financing hedging procedure. Let $P_i$ denote the terminal hedging PnL associated with the $i$-th option path. The corresponding hedging error metrics are computed as
\begin{align}
    \mathrm{RMSE}_{\mathrm{hedge}}
    &=
    \sqrt{\frac{1}{N}\sum_{i=1}^{N}P_i^2},
    \label{eq:hedging_rmse}\\
    \mathrm{MAE}_{\mathrm{hedge}}
    &=
    \frac{1}{N}\sum_{i=1}^{N}|P_i|,
    \label{eq:hedging_mae}
\end{align}
where $N$ denotes the total number of evaluated option paths. Lower values of these metrics indicate smaller replication errors associated with the estimated hedge ratios. The pricing and hedging metrics therefore evaluate two different properties of the learned pricing function, namely the accuracy of its estimated option values and the effectiveness of its corresponding hedge ratios.

\subsection{Comparison Models}
\label{subsec:comparison_models}

The proposed HARP-Net is evaluated against diverse classes of benchmark models, comprising classical, machine learning, and deep learning architectures. These models employ different mechanisms for representing nonlinear relationships and temporal dependencies among financial variables. The benchmark models utilized for comparative analysis are as follows:

\begin{enumerate}

    \item \textbf{Black--Scholes:} The Black--Scholes model \cite{black1973pricing} provides an analytical framework for pricing European options and estimating their corresponding hedge ratios.

    \item \textbf{Heston:} The Heston model \cite{heston1993closed} incorporates stochastic volatility into the underlying asset dynamics to represent time-varying market volatility.

    \item \textbf{Merton:} The Merton jump-diffusion model \cite{merton1976option} incorporates discontinuous price movements through a jump component in the underlying asset dynamics.

    \item \textbf{FEM:} The finite element method (FEM) \cite{andalaft2011real} employs numerical discretization to approximate the solution of option pricing equations.

    \item \textbf{MLP:} The multilayer perceptron (MLP) employs multiple fully connected layers with nonlinear activation functions to approximate relationships among option pricing variables \cite{hutchinson1994nonparametric}.

    \item \textbf{DLinear:} DLinear decomposes the input sequence into trend and seasonal components and independently transforms these components using linear layers \cite{zeng2023transformers}.

    \item \textbf{NLinear:} NLinear normalizes the input sequence using its final observation before applying a linear transformation, accounting for changes in the level of sequential observations \cite{zeng2023transformers}.

    \item \textbf{ResNet:} The residual neural network (ResNet) employs residual connections between successive layers to facilitate nonlinear feature learning \cite{gan2024option}.

    \item \textbf{LSTM:} The long short-term memory (LSTM) network utilizes gated recurrent operations and memory cells to capture sequential dependencies \cite{hochreiter1997long}.

    \item \textbf{GRU:} The gated recurrent unit (GRU) employs update and reset gates to control the information retained from preceding observations \cite{chung2014empirical}.

    \item \textbf{CNN-1D:} The one-dimensional convolutional neural network (CNN-1D) employs convolutional operations to extract local patterns from sequential financial observations \cite{liu2023option}.

    \item \textbf{CNN-LSTM:} The CNN-LSTM architecture integrates convolutional feature extraction with recurrent learning to represent local and sequential dependencies \cite{hu2026multi}.

    \item \textbf{CNN-BiLSTM:} The CNN-BiLSTM architecture combines convolutional layers with bidirectional LSTM layers to extract local and bidirectional temporal features within the available input window \cite{sharma2025enhancing}.

    \item \textbf{Transformer:} The Transformer utilizes self-attention mechanisms to capture dependencies among input representations without recurrent operations \cite{vaswani2017attention}.

    \item \textbf{Random Forest:} Random Forest constructs an ensemble of decision trees to represent nonlinear relationships among financial variables through aggregated predictions \cite{breiman2001random,ivascu2021option}.

    \item \textbf{XGBoost:} Extreme Gradient Boosting (XGBoost) sequentially constructs decision trees to minimize a regularized learning objective \cite{chen2016xgboost}.

    \item \textbf{DeepAR:} DeepAR employs an autoregressive recurrent architecture for probabilistic time series forecasting \cite{salinas2020deepar}.

    \item \textbf{PAttn:} PAttn utilizes patch-based input representations and attention mechanisms to capture temporal dependencies \cite{tan2024language}.

    \item \textbf{TimeXer:} TimeXer incorporates endogenous and exogenous variables into the forecasting process through attention-based feature learning \cite{wang2024timexer}.

    \item \textbf{WPMixer:} WPMixer combines wavelet decomposition with patch-based mixing to extract temporal features at different resolutions \cite{murad2025wpmixer}.

\end{enumerate}

The considered benchmark models provide different representations of financial variables and temporal information, allowing the comparative analysis to examine the contribution of analytical initialization, nonlinear feature extraction, and hierarchical refinement in the proposed HARP-Net.

\subsection{Experimental Setup}
\label{subsec:experimental_setup}

A common experimental procedure is employed across the comparative models to maintain consistency in the available financial information, data partitions, and hedging conventions. To retain the estimated option prices within their theoretical intervals, a common price transformation is employed for all neural pricing models. Let
\begin{equation}
    A_t=S_t e^{-q_t\tau_t},
    \qquad
    B_t=K e^{-r_t\tau_t},
\end{equation}
where $q_t$ and $r_t$ denote the dividend yield proxy and maturity-dependent risk-free rate, respectively. The lower and upper price bounds for European call and put options are defined as
\begin{equation}
    (\underline{V}_t,\overline{V}_t)
    =
    \begin{cases}
        (\max(A_t-B_t,0),A_t), & a=C,\\
        (\max(B_t-A_t,0),B_t), & a=P,
    \end{cases}
    \label{eq:european_price_bounds}
\end{equation}
where $a\in\{C,P\}$ represents the option type. A smooth positive-part approximation is employed instead of hard clipping as
\begin{equation}
    g_{\varepsilon}(z)
    =
    \frac{z+\sqrt{z^2+\varepsilon^2}}{2},
    \qquad \varepsilon>0.
    \label{eq:smooth_positive_part}
\end{equation}
Let $v_t$ denote the unconstrained option price obtained after inverse normalization. The corresponding bounded price is computed as
\begin{equation}
\begin{aligned}
    \widehat{V}_t
    =\underline{V}_t
    &+g_{\varepsilon}(v_t-\underline{V}_t)-g_{\varepsilon}(v_t-\overline{V}_t).
\end{aligned}
    \label{eq:smooth_price_constraint}
\end{equation}
The transformation parameter is fixed at $\varepsilon=0.1$ index points across all neural models and datasets. The bounded price is subsequently transformed into the normalized output space. This transformation is employed during training, validation, and testing, while the corresponding hedge ratio is obtained by differentiating the bounded pricing function, including its underlying-dependent price bounds. The common transformation maintains the specified price intervals without imposing additional conditions on monotonicity, convexity, or put--call parity.

All neural models are optimized using the squared terminal delta-hedging PnL objective described in Section~\ref{subsec:hedging_training}. Let $\mathcal{B}_j$ denote a mini-batch consisting of complete option paths. The corresponding training objective is expressed as
\begin{equation}
    \mathcal{L}_j(\Theta)
    =
    \frac{1}{|\mathcal{B}_j|}
    \sum_{p\in\mathcal{B}_j}P_{\Theta}(p)^2,
    \label{eq:experimental_training_loss}
\end{equation}
where $P_{\Theta}(p)$ denotes the terminal hedging PnL obtained along path $p$ using model parameters $\Theta$. The training paths are randomly shuffled at each epoch while preserving the chronological ordering of observations within each path. Model-implied hedge ratios are obtained through automatic differentiation, with the corresponding computational graph retained during training. The Adam optimizer is employed for parameter learning with a maximum global gradient norm of $5$. Training is terminated when non-finite losses or parameter gradients are encountered. The training and pathwise hedging evaluation follow the procedures described in Algorithms~\ref{alg:hedge_loss_training} and~\ref{alg:pathwise_evaluation}, respectively.

The hedging portfolio is initialized using the observed option premium, while interest accrual, dividend income, rebalancing, and terminal settlement follow the common self-financing procedure. The underlying position remains unchanged between successive available observations. The primary experiments consider delta hedging without transaction costs or additional option instruments for gamma and vega hedging. The performance under alternative market and execution conditions is examined separately in the market-realism experiments.

Hyperparameter optimization is performed independently for each neural model, dataset, and random seed using the tree-structured Parzen estimator (TPE) implemented in Hyperopt. The search considers model-specific architectural parameters, learning rate, and the number of complete option paths per mini-batch. A total of $25$ trials is allocated to each search, with a maximum of $40$ training epochs per trial and an early-stopping patience of $3$ epochs. The performance of each configuration is evaluated after every epoch using the complete validation partition. The model-specific hyperparameter search spaces are presented in Table~\ref{tab:neural-hyperparameter-space}. Learning rates are sampled on a logarithmic scale, dropout probabilities are sampled uniformly, and discrete architectural parameters are selected from predefined candidate sets. Although a common search budget is employed across neural models, the computational requirements may vary according to their architectural complexity and number of completed training epochs.

\begingroup
\footnotesize
\setlength{\tabcolsep}{5pt}
\renewcommand{\arraystretch}{1.05}

\begin{longtable}{
    @{}
    >{\raggedright\arraybackslash}p{0.25\linewidth}
    >{\raggedright\arraybackslash}p{0.34\linewidth}
    >{\raggedright\arraybackslash}p{0.34\linewidth}
    @{}
}

\caption{Hyperparameter search spaces of the proposed and compared models.}
\label{tab:neural-hyperparameter-space}\\

\toprule
\textbf{Models} & \textbf{Hyperparameter}
& \textbf{Search space}\\
\midrule
\endfirsthead

\toprule
\textbf{Model(s)} & \textbf{Hyperparameter}
& \textbf{Search space}\\
\midrule
\endhead

\endfoot

\bottomrule
\endlastfoot

\multirow{2}{=}{MLP}
& Hidden dimension & $\{8,16,32,64\}$ \\*
& Number of layers & $\{1,2,3,4\}$ \\
\midrule

\multirow{3}{=}{LSTM/GRU/DeepAR}
& Hidden dimension & $\{32,64,96\}$ \\*
& Number of layers & $\{1,2\}$ \\*
& Dropout & $[0,0.3]$ \\
\midrule

\multirow{5}{=}{Transformer}
& Model dimension & $\{32,64,96\}$ \\*
& Attention heads & $\{2,4,8\}$ \\*
& Encoder layers & $\{1,2,3\}$ \\*
& Feed-forward dimension & $\{128,192,256\}$ \\*
& Dropout & $[0,0.2]$ \\
\midrule

\multirow{3}{=}{ResNet}
& Channels & $\{16,32,64\}$ \\*
& Residual blocks & $\{2,3,4\}$ \\*
& Kernel size & $\{3,5\}$ \\
\midrule

\multirow{2}{=}{DLinear}
& Individual channel head
& $\{\mathrm{False},\mathrm{True}\}$ \\*
& Moving-average kernel
& $\{3,4,5,8,10,12,16,25\}$ \\
\midrule

\multirow{2}{=}{NLinear}
& Individual channel head
& $\{\mathrm{False},\mathrm{True}\}$ \\*
& Channel dependence & $\{2,4,8,16\}$ \\
\midrule

\multirow{4}{=}{CNN-1D}
& Number of channels & $\{32,64,96\}$ \\*
& Number of layers & $\{2,3,4\}$ \\*
& Kernel size & $\{3,5\}$ \\*
& Dropout & $[0,0.3]$ \\
\midrule

\multirow{5}{=}{\makecell[l]{CNN-LSTM/\\CNN-BiLSTM}}
& Convolutional channels & $\{16,32,64\}$ \\*
& LSTM hidden dimension & $\{32,64,96\}$ \\*
& LSTM layers & $\{1,2\}$ \\*
& Dropout & $[0,0.3]$ \\*
& Kernel size (CNN-BiLSTM) & $\{3,5\}$ \\
\midrule

\multirow{7}{=}{PAttn}
& Model dimension & $\{32,64,96\}$ \\*
& Feed-forward dimension & $\{64,128,192\}$ \\*
& Attention heads & $\{2,4,8\}$ \\*
& Dropout & $[0,0.2]$ \\*
& Prediction head dimension & $\{16,32,64\}$ \\*
& Patch length & $\{4,8,10\}$ \\*
& Stride & $\{2,4\}$ \\
\midrule

\multirow{7}{=}{TimeXer}
& Model dimension & $\{32,64,96\}$ \\*
& Feed-forward dimension & $\{64,128,192\}$ \\*
& Attention heads & $\{2,4,8\}$ \\*
& Dropout & $[0,0.2]$ \\*
& Prediction head dimension & $\{16,32,64\}$ \\*
& Patch length & $\{2,4,5,10\}$ \\*
& Encoder layers & $\{1,2\}$ \\
\midrule

\multirow{8}{=}{WPMixer}
& Model dimension & $\{32,64,96\}$ \\*
& Feed-forward dimension & $\{64,128,192\}$ \\*
& Attention heads & $\{2,4,8\}$ \\*
& Dropout & $[0,0.2]$ \\*
& Prediction head dimension & $\{16,32,64\}$ \\*
& Stride & $\{2,4\}$ \\*
& Patch length & $\{2,4,5,10\}$ \\*
& Temporal / Feature factor & $\{3,5\}$ \\
\midrule

\multirow{10}{=}{\textbf{HARP-Net}}
& Latent dimension ($d$) & $\{32,64,96\}$ \\*
& Refinement steps ($J$) & $\{2,4\}$ \\*
& Context layers & $\{1,2,3\}$ \\*
& Field layers ($H_f$) & $\{1,2,3\}$ \\*
& Activation function
& $\{\mathrm{GELU},\mathrm{ReLU},\mathrm{SiLU}\}$ \\*
& Dropout & $[0,0.15]$ \\*
& Residual scale ($\lambda_r$)
& $\{0.05,0.10,0.25,0.50\}$ \\*
& Prior correction scale ($\lambda_{\mathrm{cp}}$)
& $\{0.25,0.50,1.00\}$ \\*
& Drift clipping threshold ($c_d$)
& $\{0.02,0.05,0.10\}$ \\

\end{longtable}

\par
\endgroup

The experiments involving neural models are repeated independently using five random seeds $\mathcal{S}=\{40,41,42,43,44\}$. Hyperparameter optimization and final model training are performed separately for each seed while retaining identical contract partitions. The reported evaluation metrics are averaged across the five independent runs, whereas deterministic classical models are evaluated once for each dataset. The experiments are implemented in Python using PyTorch 2.4.1 and NumPy 1.26.4, with Hyperopt employed for hyperparameter optimization. The primary neural model comparisons are conducted on a CPU.

\section{Results and discussion}
\label{sec:results}

This section presents the empirical results obtained by comparing the proposed model with various existing approaches. Three assessment metrics are utilized to evaluate the hedging performance, followed by statistical tests to examine the reliability of the comparative results. Furthermore, the study examines the contribution of the individual components of the proposed model, the behavior of the Greeks, and the performance under varying market conditions. Finally, the computational efficiency and robustness of the models are discussed.

\subsection{Comparative hedging performance}
\label{subsec:main_hedging_results}

The results reported in Table~\ref{tab:hedging_summary} provide a comparison of the proposed HARP-Net with compared baselines across the four option datasets. The proposed model consistently achieves the lowest PnL$^2$ and Std PnL across the datasets, indicating improved performance. Among the classical approaches, Black--Scholes performs competitively, particularly on the DJX Put, where the improvement achieved with HARP-Net is relatively small. The binomial model also exhibits competitive performance on the DJX datasets in the full comparative study. However, the improvements achieved with the proposed model are more pronounced on the XEO datasets than with classical approaches.

The comparative results against deep learning models further demonstrate the effectiveness of incorporating analytical pricing information into the learning framework. Among the alternatives, GRU, LSTM, and ResNet perform competitively on specific datasets, while the proposed model maintains lower squared hedging errors across both indices and option types. Notably, models such as DeepAR, ResNet, TimeXer, and Transformer yield lower absolute Mean PnL on individual datasets among the models reported in Table~\ref{tab:hedging_summary}. However, these models exhibit greater dispersion in the terminal hedging errors. The proposed HARP-Net achieves the best performance in eight of the twelve assessment criteria, highlighting the contribution of the analytical prior and the learned corrections in reducing the variability of replication errors. The statistical significance of these improvements is examined in the subsequent analysis.

\begin{table}[!htbp]
\centering
\caption{Comparison results using PnL$^2$, Mean PnL, and Std PnL.}
\label{tab:hedging_summary}

\footnotesize
\setlength{\tabcolsep}{4.5pt}
\renewcommand{\arraystretch}{1.02}

\begin{adjustbox}{max width=\linewidth}
\begin{tabular}{@{}lrrrrrr@{}}

\toprule
& \multicolumn{3}{c}{\textbf{Call}}
& \multicolumn{3}{c}{\textbf{Put}} \\
\cmidrule(lr){2-4}
\cmidrule(lr){5-7}
\textbf{Model}
& \textbf{PnL$^2$}
& \textbf{Mean PnL}
& \textbf{Std PnL}
& \textbf{PnL$^2$}
& \textbf{Mean PnL}
& \textbf{Std PnL} \\
\midrule

\multicolumn{7}{@{}l}{\textbf{DJX}} \\
\addlinespace[2pt]

Black-Scholes
& 9.4984 & -0.7874 & 2.9797
& 9.3765 & -0.8646 & 2.9375 \\

Merton
& 9.9366 & -0.7809 & 3.0540
& 9.8079 & -0.8616 & 3.0109 \\

MLP
& 80.7721 & -1.6871 & 8.8188
& 102.7718 & -1.7754 & 9.9059 \\

NLinear
& 75.9228 & -1.7265 & 8.5232
& 58.6058 & -1.2217 & 7.4963 \\

ResNet
& 66.5041 & -1.5469 & 7.7902
& 41.3395 & \best{-0.3714} & 6.1887 \\

LSTM
& 58.0720 & -1.0499 & 7.3199
& 42.2854 & -1.0595 & 5.9823 \\

GRU
& 55.4906 & -1.3939 & 7.1615
& 42.2921 & -1.6587 & 6.0669 \\

Transformer
& 37.5509 & -0.8191 & 5.8235
& 45.7768 & -1.1252 & 6.4580 \\

DeepAR
& 43.6272 & \best{-0.7718} & 6.2450
& 46.2277 & -1.1945 & 6.3409 \\

PAttn
& 55.2659 & -1.2863 & 7.3071
& 54.5771 & -1.4110 & 7.1153 \\

TimeXer
& 60.0527 & -1.2101 & 7.5974
& 51.7078 & -1.3748 & 6.9461 \\

WPMixer
& 54.3657 & -1.0177 & 7.2160
& 53.6270 & -1.5256 & 7.0999 \\

\addlinespace[2pt]
\textbf{HARP-Net}
& \best{8.0579} & -0.7912 & \best{2.7250}
& \best{9.1811} & -0.8634 & \best{2.9042} \\

\midrule

\multicolumn{7}{@{}l}{\textbf{XEO}} \\
\addlinespace[2pt]

Black-Scholes
& 1046.2901 & -8.2090 & 31.2874
& 1194.0007 & -9.7260 & 33.1573 \\

Merton
& 1058.4611 & -8.2297 & 31.4759
& 1215.0222 & -9.7628 & 33.4621 \\

MLP
& 2689.8095 & -8.6019 & 50.9451
& 3512.0445 & -11.2817 & 57.9775 \\

NLinear
& 2758.1223 & -13.2186 & 50.3581
& 2217.4424 & -9.4013 & 45.8274 \\

ResNet
& 2483.4622 & -9.7606 & 48.5940
& 1685.8034 & -5.5755 & 40.1646 \\

LSTM
& 1416.9310 & -8.5025 & 36.3481
& 1458.4570 & -8.2446 & 36.9364 \\

GRU
& 1407.8501 & -9.0139 & 36.2082
& 1491.2501 & -8.1293 & 37.6212 \\

Transformer
& 1596.3804 & -9.6909 & 38.5834
& 1548.4253 & \best{-4.6103} & 37.2539 \\

DeepAR
& 1620.7586 & -9.1171 & 38.8146
& 1635.4944 & -9.0583 & 39.0813 \\

PAttn
& 3072.2558 & -13.1016 & 53.1802
& 3135.0199 & -10.6805 & 54.6083 \\

TimeXer
& 1747.7760 & \best{-2.2393} & 40.5718
& 3136.5397 & -13.7570 & 54.0798 \\

WPMixer
& 2528.2222 & -11.4311 & 48.6654
& 2832.4436 & -9.1147 & 52.0679 \\

\addlinespace[2pt]
\textbf{HARP-Net}
& \best{828.3690} & -8.1530 & \best{27.5955}
& \best{1070.6956} & -9.7991 & \best{31.2191} \\

\bottomrule
\end{tabular}
\end{adjustbox}

\end{table}

\subsection{Statistical significance}
\label{subsec:statistical-comparison}

The research also investigates the statistical significance of the observed differences in hedging performance. The Model Confidence Set procedure is utilized to identify the models that exhibit statistically competitive performance across the assessment metrics.

\subsubsection{Model confidence set}
\label{subsec:mcs-test}

The Model Confidence Set (MCS) procedure is employed to assess the relative performance of multiple models by sequentially eliminating models that exhibit inferior performance. The null hypothesis states that the models in the current set exhibit equal expected loss. If the hypothesis is rejected, an inferior model is eliminated, and the procedure is repeated. The resulting set consists of models whose inferior performance has not been established at the specified significance level.

Table~\ref{tab:mcs_test} presents the results of the MCS procedure, where the bold values indicate the inclusion of the corresponding models in the confidence set at the 5\% significance level. The proposed HARP-Net is retained across all the datasets and assessment metrics. Notably, Black--Scholes and the binomial model also exhibit competitive performance with respect to PnL$^2$. The proposed model is the only model retained for Std PnL on both DJX datasets, while several classical approaches are also retained on XEO. These findings support the consistency of the proposed model across the metrics, although the statistical superiority of HARP-Net over all the comparison models cannot be established.

\begingroup\footnotesize
\setlength{\tabcolsep}{3pt}\renewcommand{\arraystretch}{0.96}
\begin{longtable}{@{}llrrrrrr@{}}
\caption{Model Confidence Set $p$-value results.}
    \label{tab:mcs_test} \\
\toprule
\textbf{Model} & \textbf{Index} & \multicolumn{2}{c}{\textbf{PnL$^2$}} & \multicolumn{2}{c}{\textbf{$|\mathrm{Mean\ PnL}|$}} & \multicolumn{2}{c}{\textbf{Std PnL}} \\
\cmidrule(lr){3-4}\cmidrule(lr){5-6}\cmidrule(lr){7-8}
 & & \textbf{C} & \textbf{P} & \textbf{C} & \textbf{P} & \textbf{C} & \textbf{P} \\
\midrule\endfirsthead
\toprule
\textbf{Model} & \textbf{Index} & \multicolumn{2}{c}{\textbf{PnL$^2$}} & \multicolumn{2}{c}{\textbf{$|\mathrm{Mean\ PnL}|$}} & \multicolumn{2}{c}{\textbf{Std PnL}} \\
\cmidrule(lr){3-4}\cmidrule(lr){5-6}\cmidrule(lr){7-8}
 & & \textbf{C} & \textbf{P} & \textbf{C} & \textbf{P} & \textbf{C} & \textbf{P} \\
\midrule\endhead
\bottomrule\endfoot
Black-Scholes & DJX & \best{0.080} & \best{0.071} & $<0.001$ & 0.002 & 0.046 & 0.007 \\*
 & XEO & \best{0.120} & \best{0.139} & $<0.001$ & 0.005 & \best{1.000} & \best{0.182} \\
Heston & DJX & \best{0.073} & 0.041 & $<0.001$ & $<0.001$ & $<0.001$ & $<0.001$ \\*
 & XEO & 0.021 & \best{0.120} & $<0.001$ & $<0.001$ & $<0.001$ & 0.017 \\
Merton & DJX & 0.005 & 0.009 & $<0.001$ & $<0.001$ & $<0.001$ & $<0.001$ \\*
 & XEO & $<0.001$ & \best{0.126} & $<0.001$ & $<0.001$ & \best{0.372} & \best{0.182} \\
Binomial & DJX & \best{0.080} & \best{0.071} & $<0.001$ & 0.044 & 0.046 & $<0.001$ \\*
 & XEO & \best{0.120} & \best{0.139} & $<0.001$ & 0.004 & \best{0.643} & \best{0.182} \\
FDM & DJX & 0.001 & 0.011 & \best{0.207} & \best{1.000} & $<0.001$ & $<0.001$ \\*
 & XEO & \best{0.087} & \best{0.139} & $<0.001$ & $<0.001$ & 0.006 & $<0.001$ \\
MLP & DJX & $<0.001$ & 0.019 & \best{0.207} & $<0.001$ & $<0.001$ & $<0.001$ \\*
 & XEO & 0.047 & 0.003 & 0.007 & $<0.001$ & $<0.001$ & $<0.001$ \\
DLinear & DJX & \best{0.080} & \best{0.071} & \best{0.146} & 0.003 & $<0.001$ & $<0.001$ \\*
 & XEO & \best{0.070} & \best{0.126} & 0.014 & $<0.001$ & $<0.001$ & $<0.001$ \\
NLinear & DJX & \best{0.080} & 0.027 & 0.030 & 0.003 & $<0.001$ & $<0.001$ \\*
 & XEO & 0.047 & \best{0.073} & $<0.001$ & \best{0.526} & $<0.001$ & $<0.001$ \\
ResNet & DJX & 0.005 & \best{0.071} & \best{0.207} & \best{0.357} & $<0.001$ & $<0.001$ \\*
 & XEO & $<0.001$ & 0.015 & $<0.001$ & \best{0.504} & $<0.001$ & $<0.001$ \\
LSTM & DJX & 0.005 & \best{0.057} & \best{0.051} & 0.012 & $<0.001$ & $<0.001$ \\*
 & XEO & \best{0.070} & 0.003 & 0.014 & \best{0.124} & $<0.001$ & $<0.001$ \\
GRU & DJX & \best{0.080} & \best{0.071} & $<0.001$ & 0.037 & $<0.001$ & $<0.001$ \\*
 & XEO & \best{0.070} & 0.033 & 0.014 & \best{0.340} & $<0.001$ & $<0.001$ \\
CNN-1D & DJX & 0.018 & \best{0.071} & \best{0.207} & $<0.001$ & $<0.001$ & $<0.001$ \\*
 & XEO & $<0.001$ & 0.003 & $<0.001$ & 0.016 & $<0.001$ & $<0.001$ \\
CNN-LSTM & DJX & \best{0.080} & \best{0.071} & \best{0.063} & 0.002 & $<0.001$ & $<0.001$ \\*
 & XEO & 0.023 & \best{0.079} & $<0.001$ & \best{1.000} & $<0.001$ & $<0.001$ \\
CNN-BiLSTM & DJX & \best{0.060} & \best{0.071} & \best{0.207} & $<0.001$ & $<0.001$ & $<0.001$ \\*
 & XEO & 0.047 & 0.042 & $<0.001$ & \best{0.634} & $<0.001$ & $<0.001$ \\
Transformer & DJX & 0.018 & $<0.001$ & 0.006 & $<0.001$ & $<0.001$ & $<0.001$ \\*
 & XEO & 0.021 & $<0.001$ & 0.007 & \best{0.783} & $<0.001$ & $<0.001$ \\
DeepAR & DJX & \best{0.056} & $<0.001$ & 0.006 & 0.020 & $<0.001$ & $<0.001$ \\*
 & XEO & \best{0.068} & 0.015 & 0.001 & \best{0.129} & $<0.001$ & $<0.001$ \\
PAttn & DJX & \best{0.080} & \best{0.071} & \best{0.189} & \best{0.212} & $<0.001$ & $<0.001$ \\*
 & XEO & \best{0.070} & \best{0.139} & 0.001 & 0.048 & $<0.001$ & 0.003 \\
TimeXer & DJX & 0.018 & \best{0.071} & 0.028 & \best{0.271} & $<0.001$ & $<0.001$ \\*
 & XEO & $<0.001$ & \best{0.139} & \best{0.432} & \best{0.284} & $<0.001$ & $<0.001$ \\
WPMixer & DJX & \best{0.056} & 0.037 & 0.028 & \best{0.357} & $<0.001$ & $<0.001$ \\*
 & XEO & 0.047 & \best{0.126} & $<0.001$ & \best{0.330} & $<0.001$ & $<0.001$ \\
HARP-Net & DJX & \best{1.000} & \best{1.000} & \best{1.000} & \best{0.470} & \best{1.000} & \best{1.000} \\*
 & XEO & \best{1.000} & \best{1.000} & \best{1.000} & \best{0.783} & \best{0.643} & \best{1.000} \\
\end{longtable}

\endgroup

\subsection{Ablation study}
\label{subsec:ablation-study}

The proposed model incorporates an analytical prior along with several learning components to improve hedging performance. An ablation study is conducted to examine the individual contribution of these components. The configurations considered in the study are presented in Table~\ref{tab:ablation-djx-module-table}, where P, R, T, D, F, B, and G denote the Black--Scholes prior, residual anchor, temporal encoder, domain embeddings, current-feature skip, bridge refinement, and velocity gate, respectively. Case I represents the proposed model, Case II utilizes only Black--Scholes, and Case III integrates a current-state MLP with the analytical prior. The remaining variants are obtained by excluding individual components from the proposed architecture.

\begin{table}[!htbp]
\centering
\footnotesize
\setlength{\tabcolsep}{5pt}
\renewcommand{\arraystretch}{1.05}
\caption{Comparison results for the variants of HARP-Net.}
\label{tab:ablation-djx-module-table}
\begin{adjustbox}{max width=\linewidth}
\begin{tabular}{@{}lcccccccrr@{}}
\toprule
\textbf{Case}
& \textbf{P} & \textbf{R} & \textbf{T} & \textbf{D}
& \textbf{F} & \textbf{B} & \textbf{G}
& \multicolumn{2}{c}{\textbf{Normalized average}} \\
\cmidrule(lr){9-10}
& & & & & & & & \textbf{PnL$^2$} & \textbf{Std} \\
\midrule
I
& \checkmark & \checkmark & \checkmark & \checkmark
& \checkmark & \checkmark & \checkmark
& \textbf{1.0000} & \textbf{1.0000} \\
II
& \checkmark & $\times$ & $\times$ & $\times$
& $\times$ & $\times$ & $\times$
& 1.1016 & 1.0535 \\
III
& \checkmark & \checkmark & $\times$ & $\times$
& \checkmark & $\times$ & $\times$
& 1.0160 & 1.0062 \\
IV
& \checkmark & \checkmark & $\times$ & \checkmark
& \checkmark & \checkmark & \checkmark
& 1.0078 & 1.0041 \\
V
& \checkmark & $\times$ & \checkmark & \checkmark
& \checkmark & \checkmark & \checkmark
& 1.1028 & 1.0522 \\
VI
& \checkmark & \checkmark & \checkmark & $\times$
& \checkmark & \checkmark & \checkmark
& 1.0734 & 1.0333 \\
VII
& \checkmark & \checkmark & \checkmark & \checkmark
& $\times$ & \checkmark & \checkmark
& 1.0102 & 1.0052 \\
IX
& \checkmark & \checkmark & \checkmark & \checkmark
& \checkmark & \checkmark & $\times$
& 1.1003 & 1.0414 \\
\bottomrule
\end{tabular}
\end{adjustbox}
\end{table}

Table~\ref{tab:ablation-djx-module-table} reports the normalized errors to facilitate comparison across the variants. The proposed model consistently yields the lowest reported normalized errors. The incorporation of a current-state MLP improves the performance of the analytical model, while excluding the residual anchor or the velocity gate from HARP-Net leads to comparatively higher squared hedging errors. The removal of domain embeddings also degrades performance, highlighting the relevance of domain-specific information in learning the corrections. In contrast, the variants excluding the temporal encoder or current-feature skip exhibit relatively small variations in performance. These results emphasize the contribution of the residual anchor and velocity gate within the proposed architecture. However, the effect of removing an individual component depends on its interaction with the remaining components.

\subsection{Greek quality diagnostics}
\label{subsec:greek-quality}

The study further examines the behavior of the Greeks obtained using the proposed and comparison models. This analysis is essential as the models are trained using the hedging objective without direct supervision of the Greeks. Delta regularity, derivative magnitudes, agreement with the analytical reference, numerical stability, and attribution residuals are utilized to assess different aspects of the learned pricing functions.

\begin{figure}[!htbp]
\centering
\includegraphics[width=0.49\linewidth]{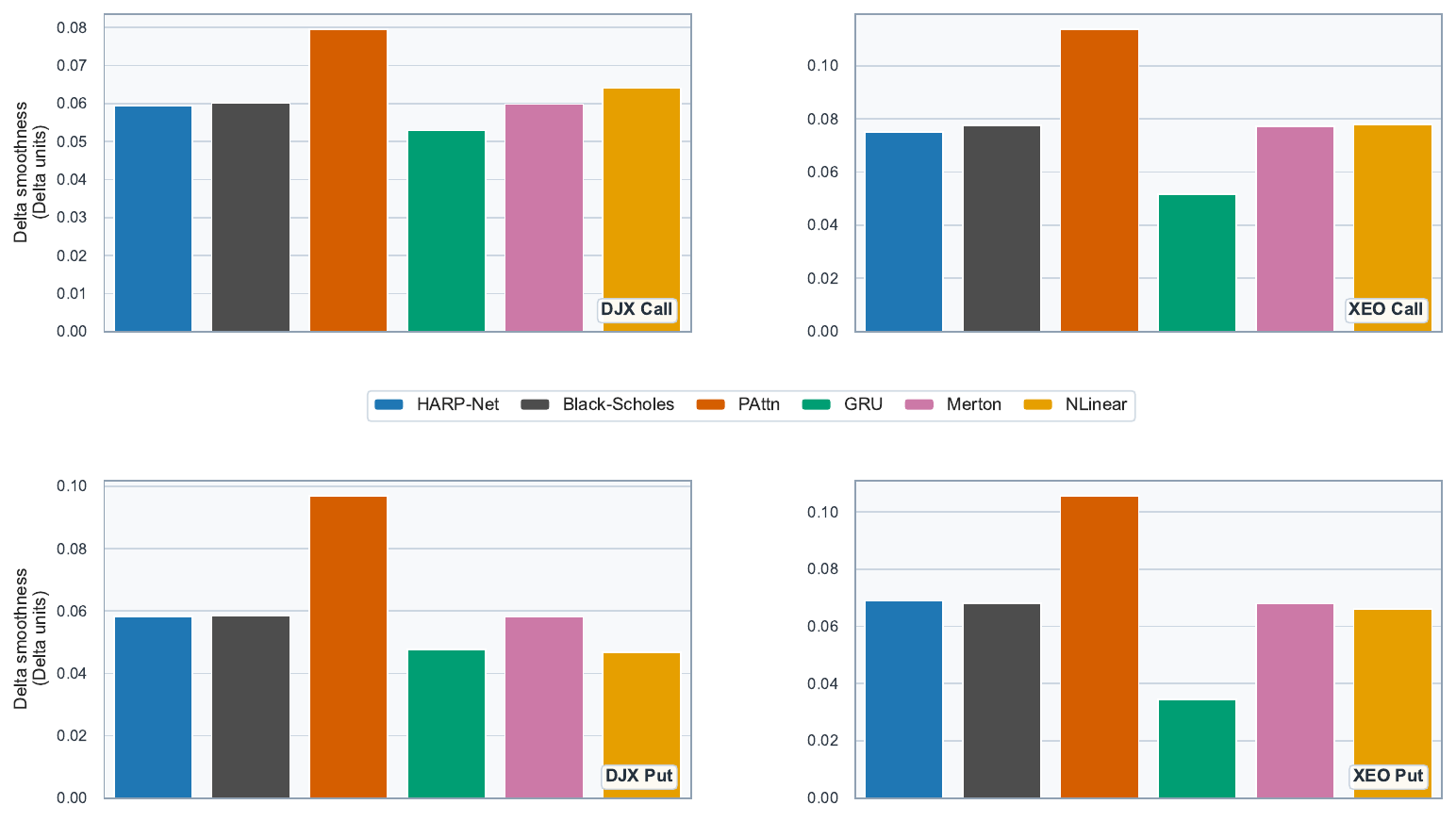}\hfill
\includegraphics[width=0.49\linewidth]{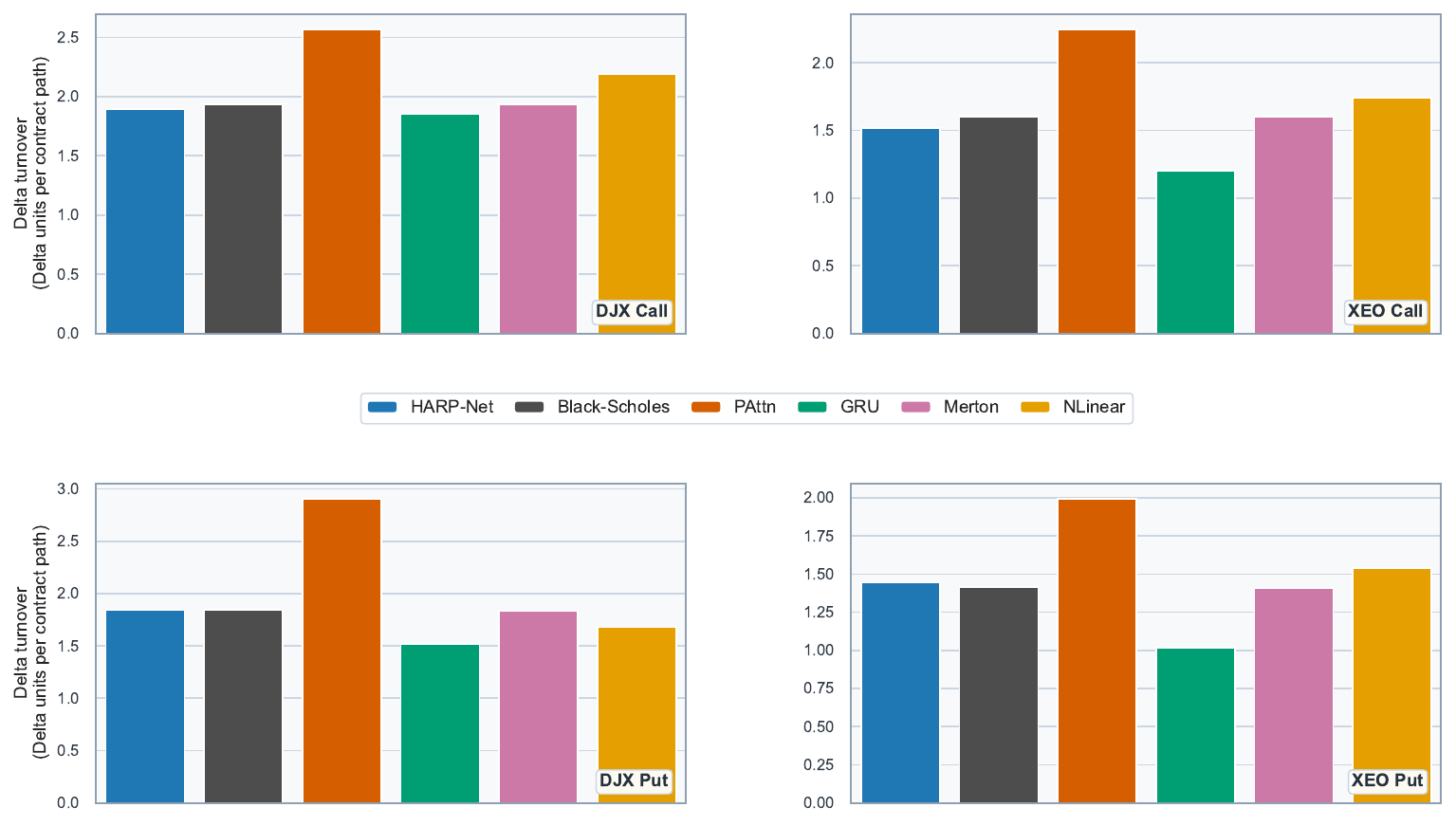}
\caption{Delta smoothness (left) and turnover (right) of the comparative models.}
\label{fig:greek-quality-pooled-overview}
\end{figure}

Fig.~\ref{fig:greek-quality-pooled-overview} illustrates the smoothness and turnover of the hedge positions across the datasets. The proposed HARP-Net exhibits behavior close to Black--Scholes, with comparatively lower turnover for the call options. Although GRU yields lower values for several configurations, the corresponding hedging errors reported in Table~\ref{tab:hedging_summary} are higher. These findings indicate that smoother hedge positions do not necessarily result in improved hedging performance and highlight the importance of retaining responsiveness to variations in market conditions.

\begin{figure}[!htbp]
\centering
\includegraphics[width=0.49\linewidth]{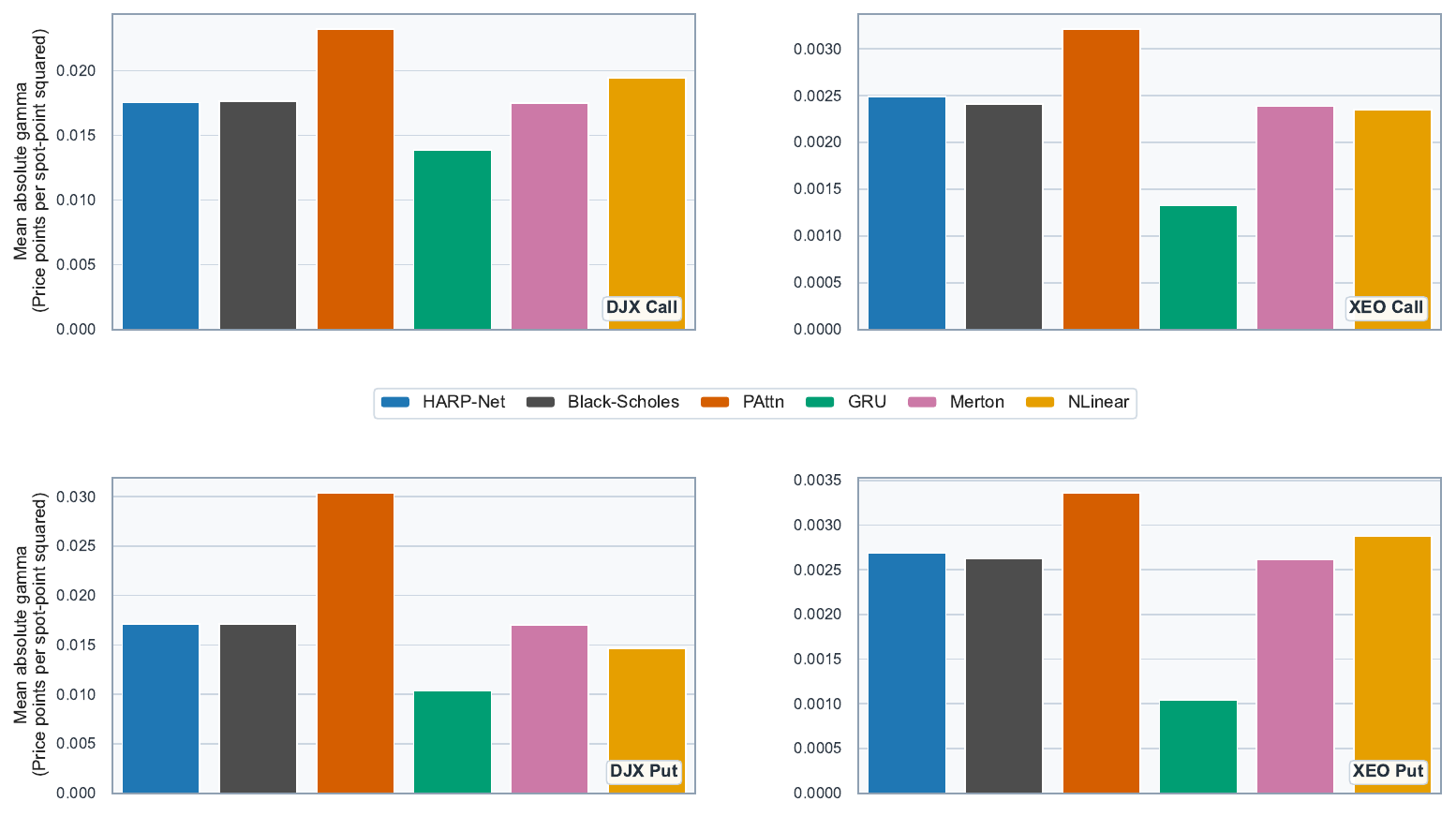}\hfill
\includegraphics[width=0.49\linewidth]{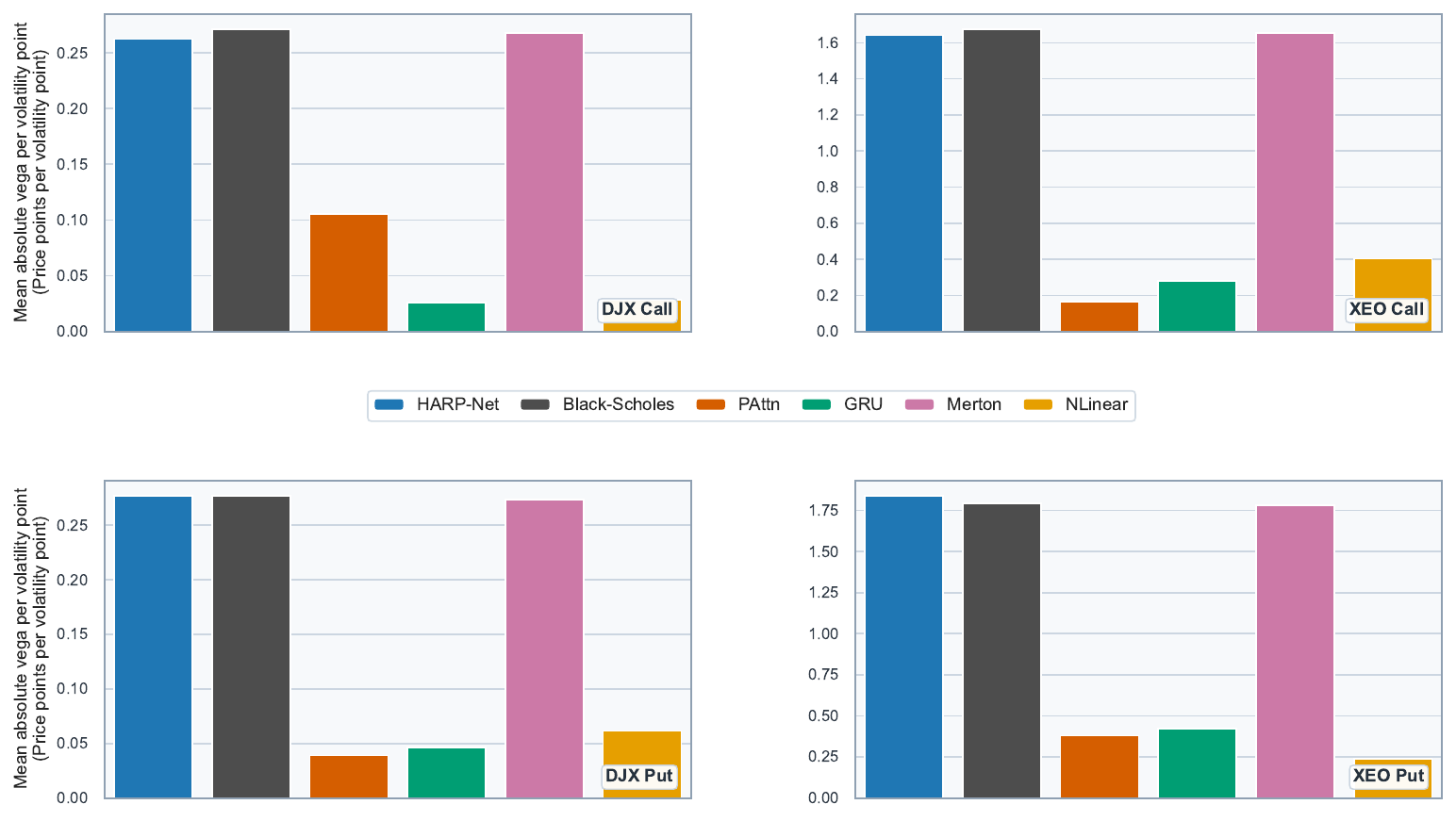}
\caption{Mean absolute gamma (left) and vega per volatility percentage point (right).}
\label{fig:greek-quality-magnitudes}
\end{figure}

The gamma and vega magnitudes obtained using the comparison models are illustrated in Fig.~\ref{fig:greek-quality-magnitudes}. HARP-Net exhibits magnitudes close to the classical models, whereas the neural alternatives exhibit greater variation, particularly with respect to vega. Furthermore, the delta obtained using HARP-Net exhibits a strong correlation with the Black--Scholes delta. The results suggest that integrating the analytical prior allows the proposed model to retain sensitivity characteristics close to those of the reference model while improving its hedging performance. However, reference agreement alone does not establish the accuracy of the Greeks.

The numerical stability of the Greeks is also examined under different finite-difference perturbations. Although none of the comparison models exhibits nonfinite derivatives or exceeds the prescribed explosion thresholds, the proposed model exhibits comparatively lower sensitivity to variations in the perturbation size. This finding indicates the stability of the estimated derivatives under the considered numerical settings.

\begin{figure}[!htbp]
\centering
\includegraphics[width=0.95\linewidth]{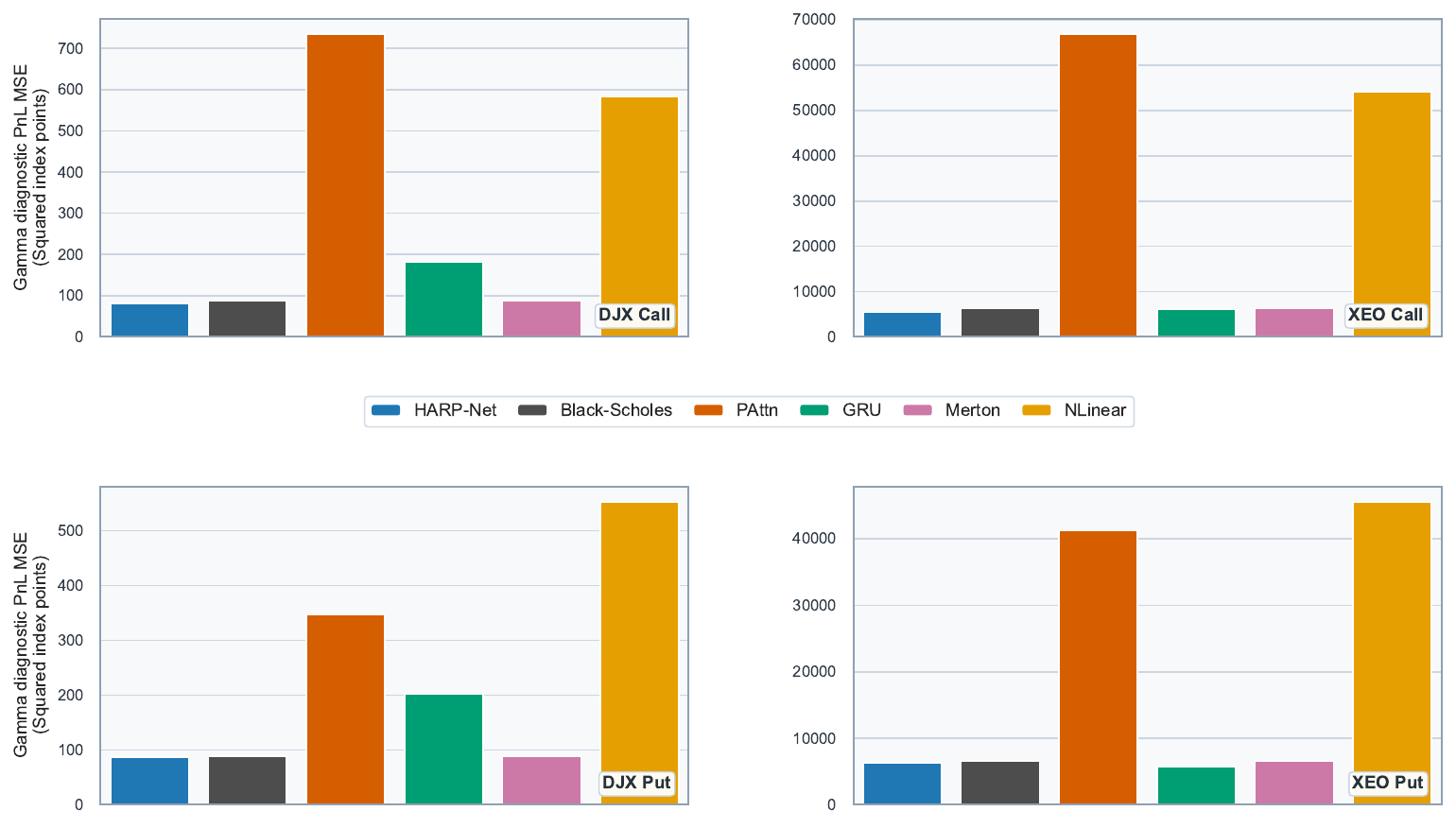}
\caption{Mean squared gamma-attribution residuals on the test datasets.}
\label{fig:greek-quality-gamma-diagnostic-grid}
\end{figure}

\begin{figure}[!htbp]
\centering
\includegraphics[width=0.95\linewidth]{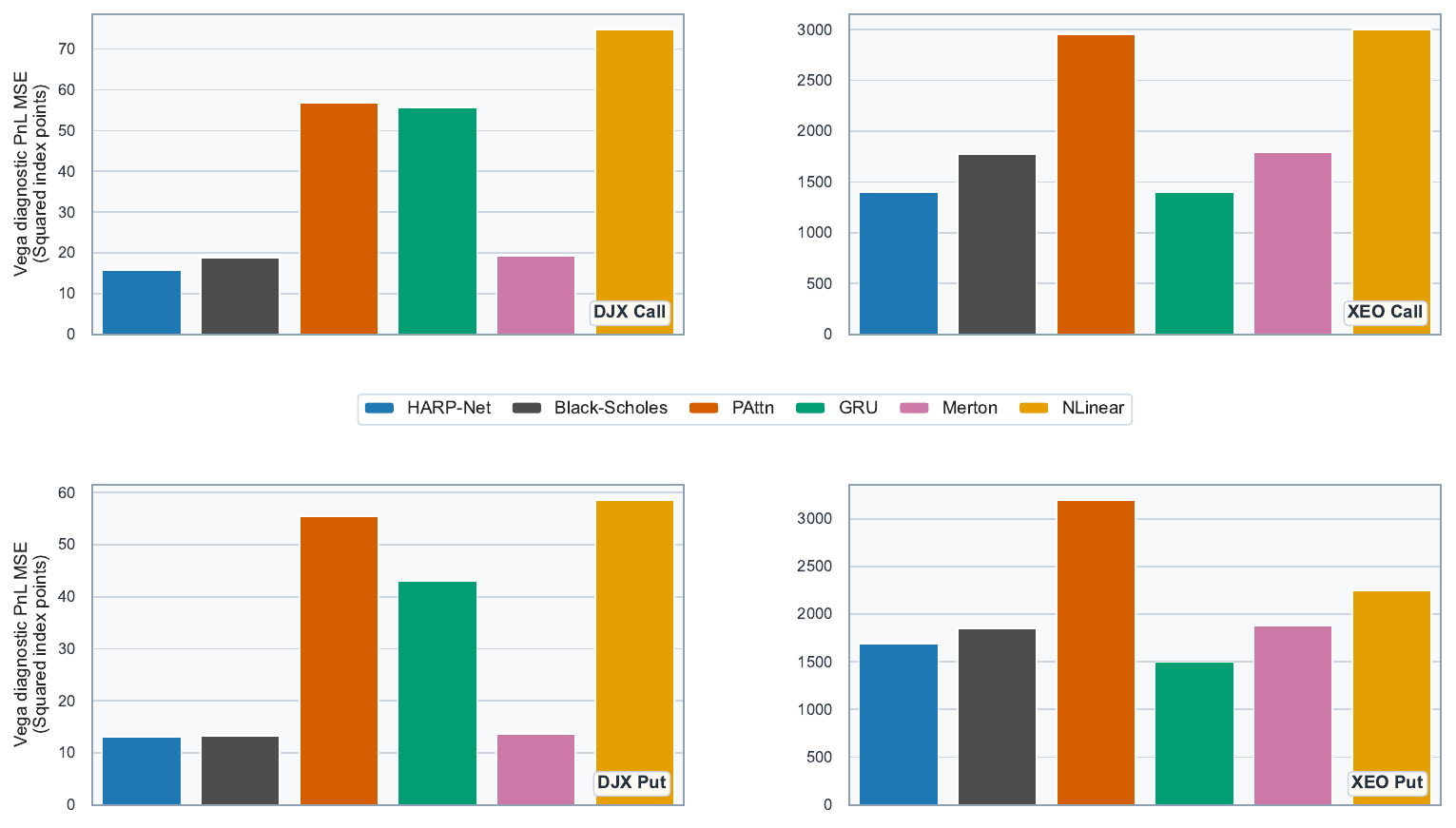}
\caption{Mean squared vega-attribution residuals on the test datasets.}
\label{fig:greek-quality-vega-diagnostic-grid}
\end{figure}

Figs.~\ref{fig:greek-quality-gamma-diagnostic-grid} and~\ref{fig:greek-quality-vega-diagnostic-grid} present the results obtained using gamma- and vega-attribution residuals, respectively. The proposed HARP-Net yields the lowest gamma residual error on both DJX datasets and XEO Call, and the lowest vega residual error on both DJX datasets. Among the alternatives, GRU performs competitively on XEO Put and achieves lower vega residual errors on the XEO datasets. These findings suggest that the improvements in delta-hedging performance are not necessarily accompanied by improvements in all derivative diagnostics. As the residuals account for only selected Taylor expansion terms, the results should not be interpreted as the performance of executable gamma- or vega-hedging strategies.

\subsection{Market-realism analysis}
\label{subsec:market-realism}

The study conducts a comparative analysis under varying execution and portfolio-management conditions to examine the robustness of the proposed model. The analysis considers different initial option proceeds, transaction costs, rebalancing frequencies, position limits, and liquidity conditions. The trained models are utilized without further training to examine their adaptability to these variations.

\begin{figure}[!htbp]
\centering
\includegraphics[width=0.95\linewidth]{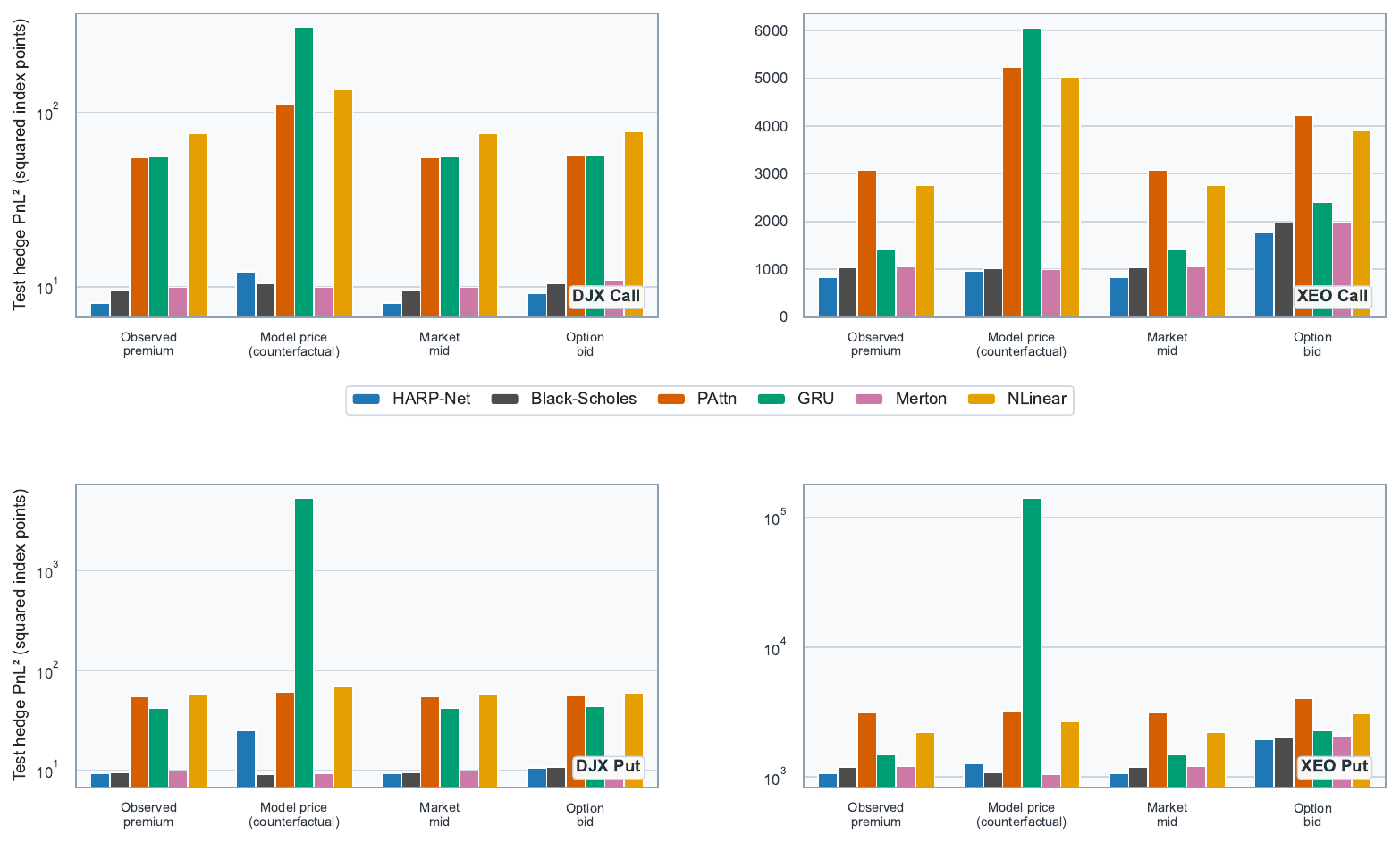}
\caption{Hedging PnL$^2$ under alternative initial option proceeds.}
\label{fig:market-realism-execution}
\end{figure}

Fig.~\ref{fig:market-realism-execution} illustrates the hedging performance under different initial option proceeds. The proposed model achieves the lowest PnL$^2$ among the selected comparison models under the observed premium, market mid-price, and option bid settings across the datasets. In contrast, the model-implied price setting exhibits greater variation among the neural models. Since the initial capital differs across models in this setting, the results reflect the combined effect of model pricing and hedge selection rather than a direct common-capital comparison.

\begin{figure}[!htbp]
\centering
\includegraphics[width=0.95\linewidth]{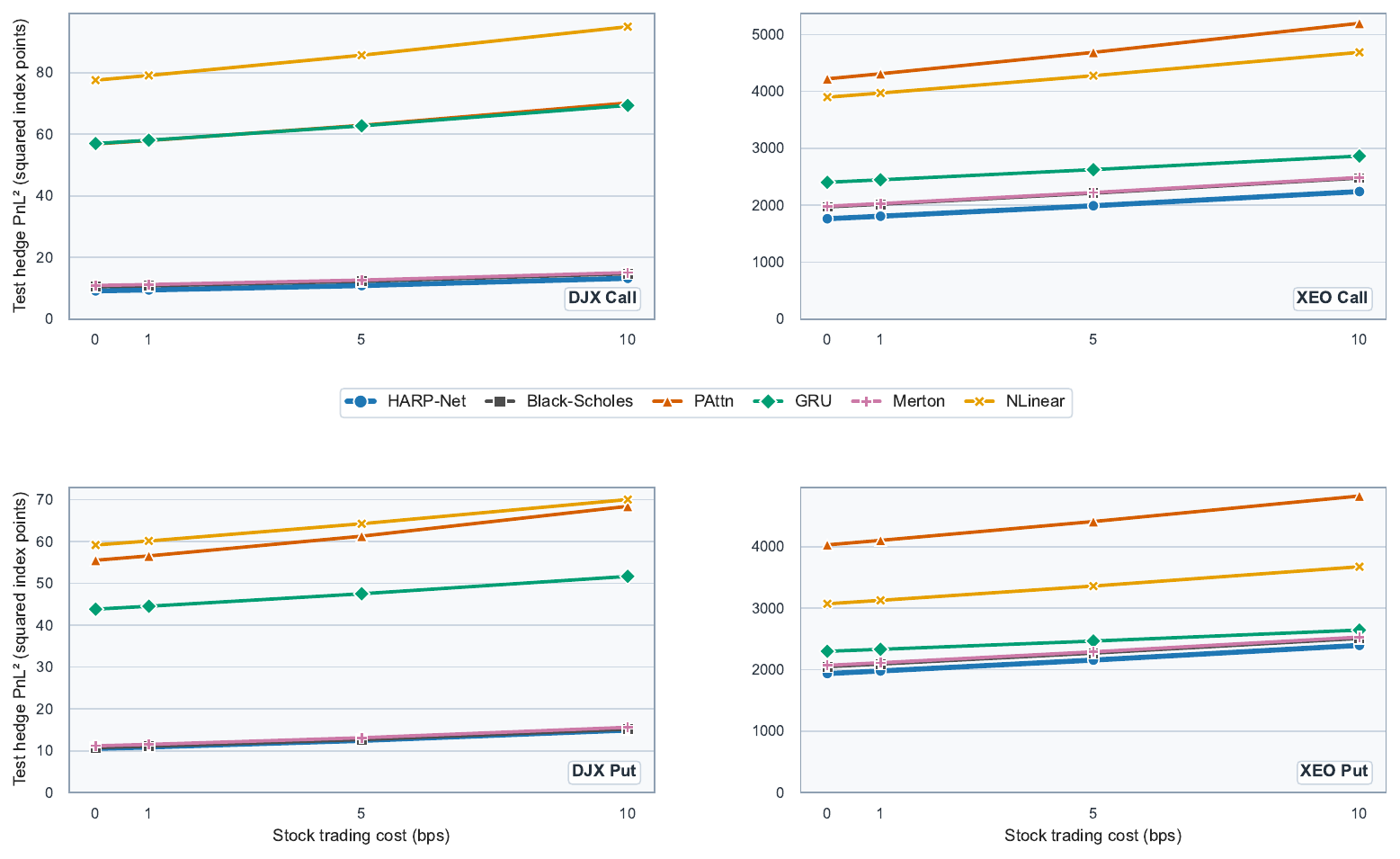}
\caption{Hedging PnL$^2$ under proportional underlying transaction costs.}
\label{fig:market-realism-transaction-costs}
\end{figure}

\begin{figure}[!htbp]
\centering
\includegraphics[width=0.95\linewidth]{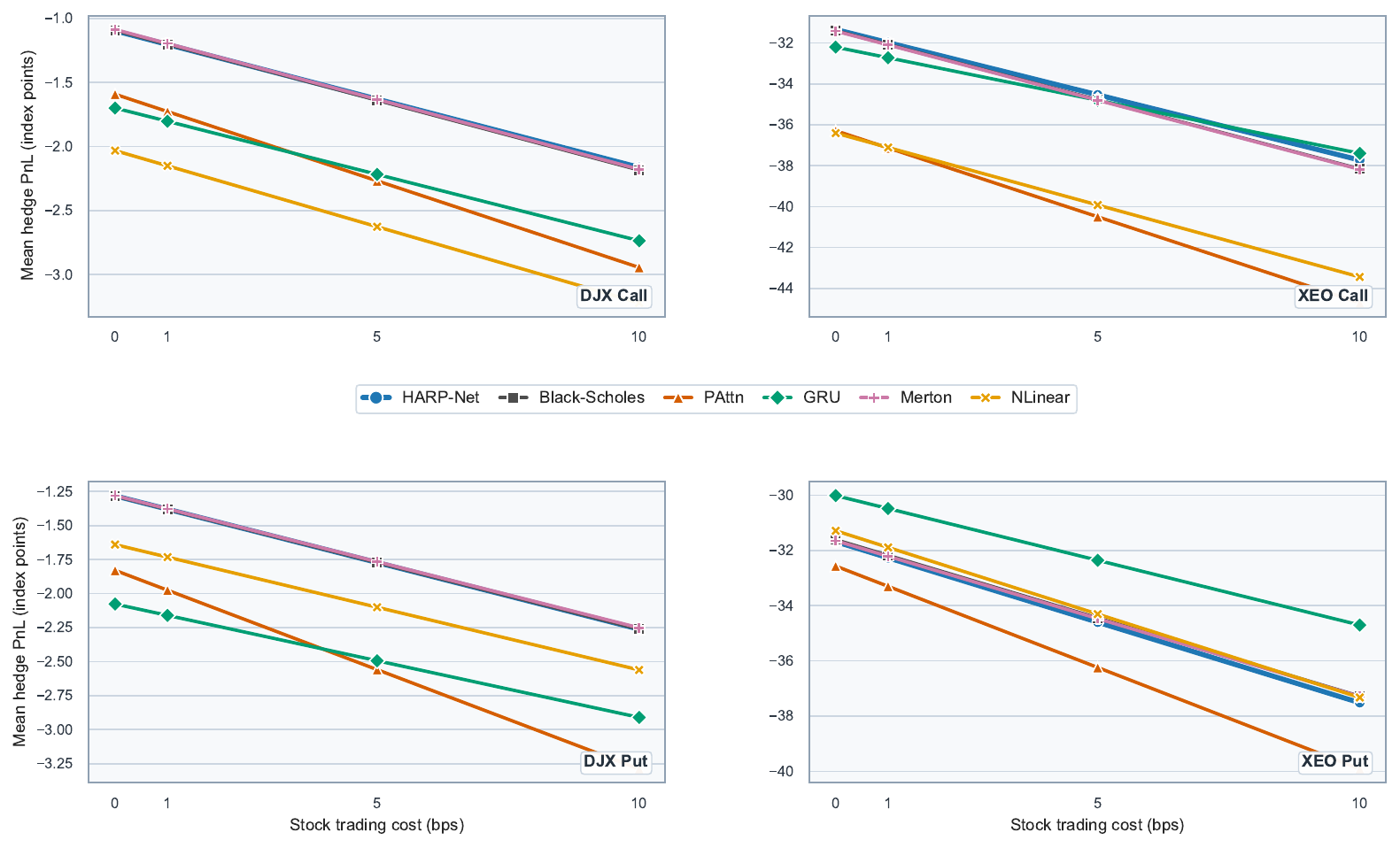}
\caption{Mean hedging PnL under proportional underlying transaction costs.}
\label{fig:market-realism-transaction-cost-mean}
\end{figure}

Figs.~\ref{fig:market-realism-transaction-costs} and~\ref{fig:market-realism-transaction-cost-mean} examine the impact of transaction costs on the hedging errors. Although the squared errors increase as transaction costs increase in the considered scenarios, the proposed HARP-Net consistently maintains the lowest PnL$^2$ among the comparison models. The corresponding Mean PnL becomes more negative as the cost increases, reflecting the expenses incurred during hedge adjustments. The consistency of HARP-Net across the cost settings indicates that its empirical performance advantage is not restricted to the absence of transaction costs.

\begin{figure}[!htbp]
\centering
\includegraphics[width=0.95\linewidth]{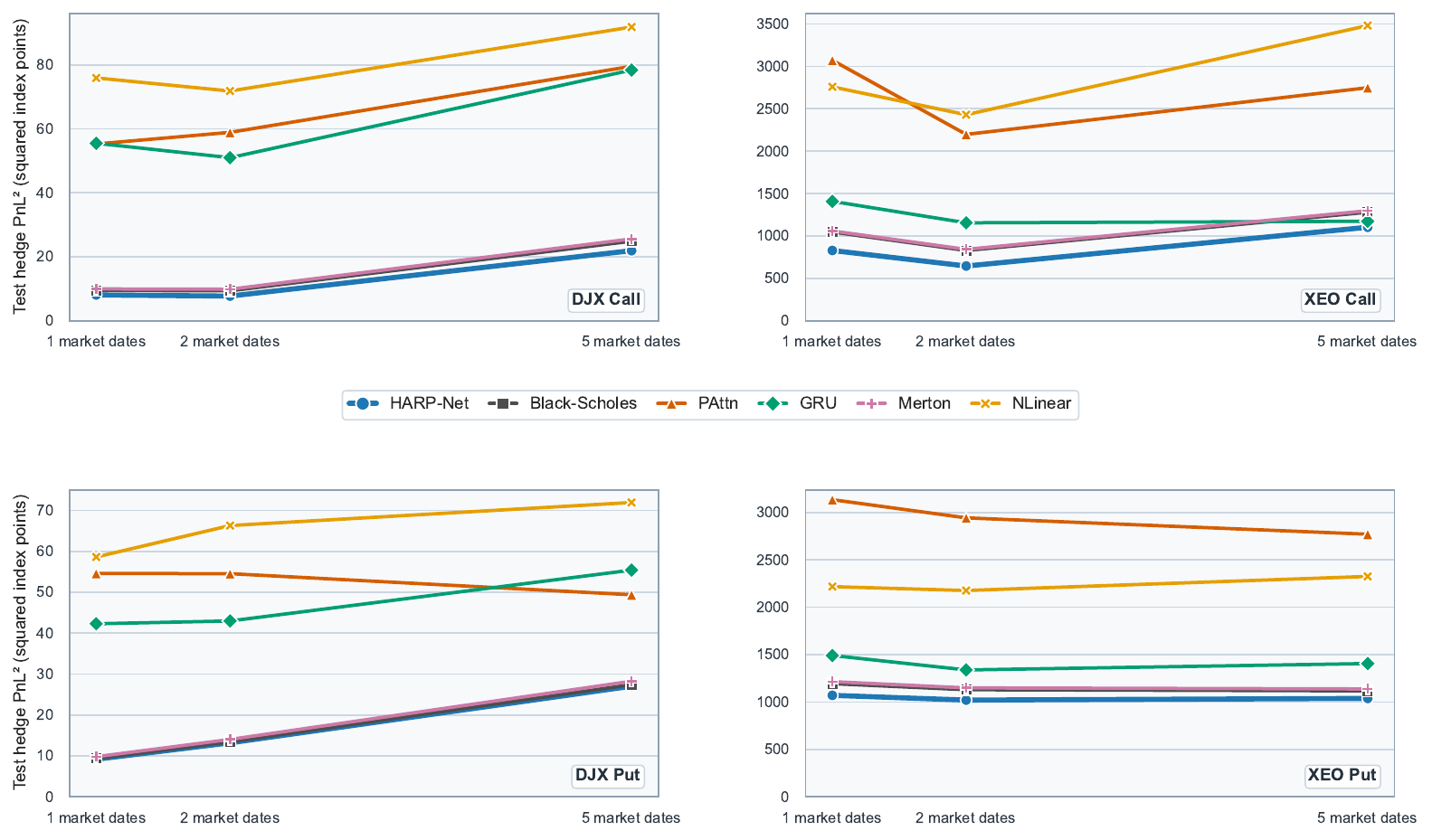}
\caption{Hedging PnL$^2$ for different rebalancing frequencies.}
\label{fig:market-realism-rebalancing}
\end{figure}

Fig.~\ref{fig:market-realism-rebalancing} presents the hedging performance obtained using varying rebalancing frequencies. The proposed HARP-Net maintains lower squared errors across the considered frequencies. Notably, reducing the frequency of rebalancing has a more pronounced adverse effect on the DJX datasets, while the XEO datasets exhibit heterogeneous behavior. This variation indicates that more frequent rebalancing does not necessarily reduce the hedging error in every dataset.

\begin{figure}[!htbp]
\centering
\includegraphics[width=0.95\linewidth]{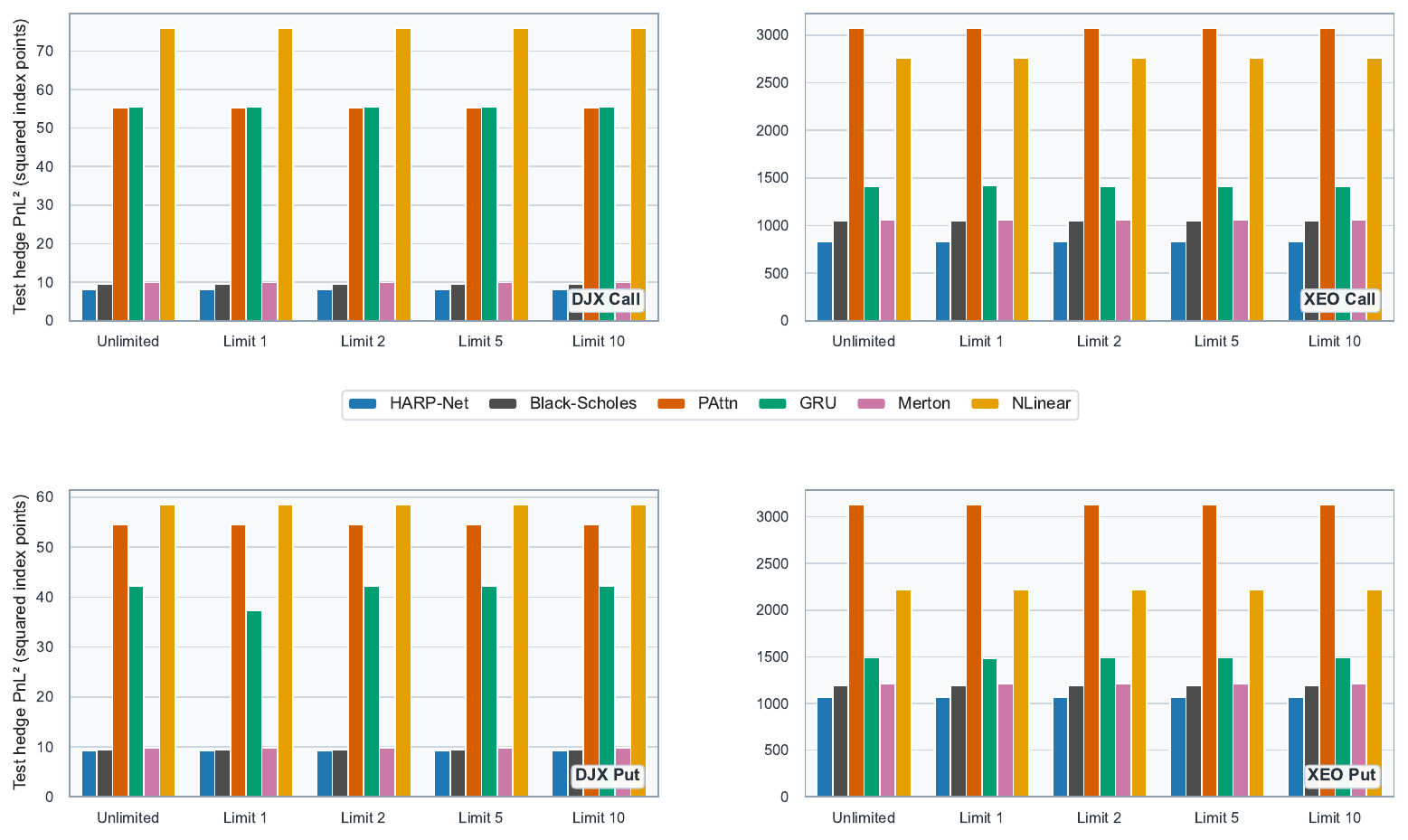}
\caption{Hedging PnL$^2$ under symmetric limits on the underlying position.}
\label{fig:market-realism-position-limits}
\end{figure}

The effect of restricting the underlying hedge positions is illustrated in Fig.~\ref{fig:market-realism-position-limits}. The proposed model maintains its performance relative to the comparison models under the considered restrictions. However, the larger position limits are nonbinding, and hence their results do not establish robustness to additional binding constraints.

\subsection{Computational efficiency and robustness}
\label{subsec:computational_efficiency}

Table~\ref{tab:computational-efficacy} presents the computational efficiency of the proposed and comparison models with respect to model complexity, training time, evaluation time, and relative hedging error. Although models such as NLinear and MLP require fewer parameters and lower computational time, their hedging errors are comparatively higher. The proposed HARP-Net achieves a balance between computational requirements and hedging performance, while the more computationally demanding classical approaches, such as Heston and FDM, exhibit competitive accuracy but higher reported evaluation times. As the classical and neural experiments are conducted separately, their runtimes should be interpreted with caution.

\begin{table}[!htbp]
\centering
\caption{Computational efficiency of the comparison models.}
\label{tab:computational-efficacy}
\small
\setlength{\tabcolsep}{6pt}
\renewcommand{\arraystretch}{1.05}
\begin{adjustbox}{max width=\linewidth}
\begin{tabular}{@{}lrrrr@{}}
\toprule
\textbf{Model} & \textbf{Parameters} & \shortstack{\textbf{Train time}\\(min/run)} & \shortstack{\textbf{Test time}\\(ms/path)} & \shortstack{\textbf{PnL$^2$}\\\textbf{ratio}} \\
\midrule
\multicolumn{5}{@{}l}{\textit{Classical models}} \\
\addlinespace[2pt]
Black-Scholes & -- & -- & 1.30 & 1.14 \\
Merton & -- & -- & 2.60 & 1.18 \\
Heston & -- & -- & 47.56 & 1.35 \\
FDM & -- & -- & 5122.51 & 1.19 \\
\midrule
\multicolumn{5}{@{}l}{\textit{Neural models}} \\
\addlinespace[2pt]
GRU & 89.1K & 10.8 & 16.19 & 2.94 \\
NLinear & 163 & 1.7 & 3.79 & 4.51 \\
PAttn & 41.9K & 12.1 & 16.37 & 4.59 \\
LSTM & 48.4K & 10.4 & 16.12 & 2.97 \\
TimeXer & 94.6K & 17.5 & 23.30 & 4.01 \\
CNN-LSTM & 68.4K & 9.4 & 18.78 & 2.48 \\
Transformer & 50.8K & 3.6 & 6.49 & 2.84 \\
WPMixer & 87.9K & 6.1 & 10.34 & 4.22 \\
CNN-BiLSTM & 185.1K & 19.7 & 34.84 & 2.44 \\
MLP & 4.3K & 1.8 & 3.57 & 5.88 \\
DLinear & 206 & 3.2 & 4.12 & 5.79 \\
DeepAR & 90.7K & 11.4 & 22.32 & 3.00 \\
ResNet & 21.8K & 5.2 & 9.94 & 3.64 \\
CNN1D & 23.8K & 4.0 & 6.92 & 5.67 \\
\midrule
\textbf{HARP-Net} & 107.1K & 5.9 & 13.21 & \textbf{1.00} \\
\bottomrule
\end{tabular}
\end{adjustbox}
\end{table}

The results reported in Table~\ref{tab:robustness-seed-pnl2} illustrate the variation in hedging performance across independent training instances. The proposed HARP-Net consistently achieves the lowest mean squared hedging error and exhibits lower variability than the selected neural models across all the datasets. These findings highlight the stability of the proposed learning framework under variations in model initialization. However, the comparison is conducted on the same test observations and does not establish stability across independent market periods.

\begin{table}[!htbp]
\centering
\caption{Mean and standard deviation of PnL$^2$ for the selected neural models.}
\label{tab:robustness-seed-pnl2}
\footnotesize
\setlength{\tabcolsep}{6pt}
\renewcommand{\arraystretch}{1.12}
\begin{adjustbox}{max width=\linewidth}
\begin{tabular}{@{}lcccc@{}}
\toprule
\textbf{Model} & \textbf{DJX Call} & \textbf{DJX Put} & \textbf{XEO Call} & \textbf{XEO Put} \\
\midrule
HARP-Net & \best{8.0579 $\pm$ 0.4946} & \best{9.1811 $\pm$ 0.2521} & \best{828.3690 $\pm$ 39.7088} & \best{1070.6956 $\pm$ 12.8283} \\
GRU & 55.4906 $\pm$ 22.5244 & 42.2921 $\pm$ 20.7488 & 1407.8501 $\pm$ 318.9742 & 1491.2501 $\pm$ 224.1459 \\
NLinear & 75.9228 $\pm$ 9.1388 & 58.6058 $\pm$ 15.8589 & 2758.1223 $\pm$ 791.3897 & 2217.4424 $\pm$ 551.1458 \\
PAttn & 55.2659 $\pm$ 5.3775 & 54.5771 $\pm$ 17.8131 & 3072.2558 $\pm$ 852.0036 & 3135.0199 $\pm$ 767.3476 \\
\bottomrule
\end{tabular}
\end{adjustbox}
\end{table}

\subsection{Discussion}
\label{subsec:overall-discussion}

The empirical findings demonstrate the effectiveness of integrating analytical pricing information with deep learning for option hedging. The improvements in squared hedging errors are primarily associated with lower dispersion of terminal replication errors, while the Greek diagnostics indicate that these improvements are obtained without substantial deviations from the analytical reference. The ablation results further highlight the importance of the residual anchor and velocity gate in learning the corrections. Furthermore, the comparative performance under varying market assumptions and training instances indicates the robustness of the proposed approach within the adopted experimental settings.

Despite these improvements, the statistical comparisons indicate that the classical approaches remain competitive on several datasets. The relatively short evaluation periods, use of synthetic underlying hedges and prescribed execution costs also limit the interpretation of the results. Further evaluation across different market periods and executable hedging instruments is necessary to assess the generalization of the proposed model.

\section{Conclusion}

In this study, we introduced HARP-Net, a hedging-aware neural option pricing network that learns an explicit and differentiable pricing surface directly aligned with dynamic hedging performance. The proposed model integrates a financially grounded analytical prior, a temporal context encoder, a domain-wise embedding strategy for heterogeneous option-state variables, a bounded-residual anchor for stable latent initialization, and a gated bridge-refinement mechanism that progressively adjusts the latent price representation toward a hedge-aware solution. By optimizing the model with a self-financing hedging objective rather than relying primarily on quote-based price regression, HARP-Net learns pricing functions whose induced model-implied deltas are better suited for practical hedge construction. The empirical results showed that HARP-Net achieves strong and consistent hedging performance across multiple datasets and option styles while maintaining a favorable balance between predictive quality and computational efficiency. Further analyses of computational efficiency, robustness across random seeds, module-wise ablation, market-realism, and the distribution of hedging errors demonstrated the contributions of the proposed components and highlighted the importance of hedging-aware learning for option pricing function modeling. Future works may extend this framework to settings with transaction costs, multi-asset hedging, and stochastic interest rates.

%%%%%%%%%%%%%%%%%%%%%%%%%%%%%%%%%%%%%%%%%%%%%%%%%%%%%%%%%%%%%%%%%%%%%   
% Bibliography
\FloatBarrier

\begingroup
\small
\setlength{\bibsep}{1pt plus 0.5pt}
\bibliographystyle{elsarticle-harv}
\bibliography{example}
\endgroup

%%%%%%%%%%%%%%%%%%%%%%%%%%%%%%%%%%%%%%%%%%%%%%%%%%%%%%%%%%%%%%%%%%%%%  

\end{document}